\documentclass[12pt]{article}

\usepackage{newtxtext,newtxmath}

\usepackage{graphicx}

\usepackage[letterpaper,margin=1in]{geometry}

\usepackage{bm}		

\renewenvironment{abstract}
	{\quotation}
	{\endquotation}

\date{}

\makeatletter
\renewcommand{\fnum@figure}{\textbf{Figure \thefigure}}
\renewcommand{\fnum@table}{\textbf{Table \thetable}}
\makeatother

\usepackage{scicite}

\usepackage{url}

\newcommand{\nn}{\nonumber \\}	
\DeclareMathOperator{\sgn}{sgn}		

\def\scititle{
	Theory of universal Planckian metal in $t$-$J$ model: application for high-$T_c$ cuprate superconductors
}
\title{\bfseries \boldmath \scititle}

\author{
	Yung-Yeh Chang$^{1,2}$,
	Khoe Van Nguyen$^{3}$,
	Kimberly Remund$^{2,3}$,
	Chung-Hou Chung$^{2,3,4\ast}$\and
	\small$^{1}$Institute of Physics, Academia Sinica, Taipei 11529, Taiwan.\and
	\small$^{2}$Physics Division, National Center for Theoretical Sciences, Taipei 10617, Taiwan.\and
	\small$^{3}$Department of Electrophysics, National Yang Ming Chiao Tung University, Hsinchu 30010, Taiwan.\and
	\small$^{4}$Center for Theoretical and Computational Physics (CTCP),\\
	\small{National Yang Ming Chiao Tung University, Hsinchu 30010, Taiwan.}\and
	\small$^\ast$Corresponding author. Email: chung0523@nycu.edu.tw
}

\begin{document} 

\maketitle


\begin{abstract} \bfseries \boldmath
The mysterious quantum-critical Planckian bad metal phase with perfect $T$-linear resistivity persisting beyond the quasi-particle limit and universal $T$-linear scattering rate $1/\tau \sim k_B T/\hbar$ has been observed in various high-$T_c$ cuprate superconductors. 
Here, we develop a realistic theoretical approach to this phase in an analytically solvable large-$N$ multi-channel Kondo lattice model, derived from a heavy-fermion formulated conventional $t$-$J$ model, known for qualitatively describing cuprates. 
This phase is originated from 
critical charge Kondo fluctuations where 
disordered local bosonic charge fluctuations couple to spinon and heavy conduction-electron Fermi surfaces near a charge-Kondo-breakdown local quantum critical point associated with pseudogap-to-Fermi liquid transition. 
Our results show excellent agreement with experiments 
and offer broad implications for other unconventional superconductors.
\end{abstract}


\textbf{Introduction$-$}
A major mystery in condensed matter systems is the microscopic origin of the strange metal state from which unconventional superconductivity directly emerges by lowering temperatures. This state has been observed in various unconventional superconductors, including: cuprate \cite{Takagi_1992_LSCO_PRL,Taillefer-planckian-2019, Ramshaw-MottPlanckian-arxiv-2024}, iron pnictides and chalcogenides \cite{Hussey-pnictide0-RPP}, 
organic \cite{Taillefer-Organic-PRB} 
and heavy-fermion compounds \cite{Custers2003Nature,Mackenzie-Science,YYC-SM-115-NatComm}, 
and twisted bi-layer graphene \cite{Cao-SM-TBG-PRL}. It is characterized by the linear-in-temperature resistivity and logarithmic-in-temperature specific heat coefficient over a wide range in temperatures. 


The most intriguing class of strange metal is the "Planckian metal" observed in cuprates, where the perfect linear-in-temperature scattering
rate reaches the maximum "Planckian dissipation limit" – bounded only by thermal energy $k_BT$ divided by the universal constant
$\hbar$ ($\hbar = h/2\pi$ with $h$ being the Planck constant): $1/\tau = \alpha k_BT/\hbar$, with $\alpha \sim 1$ a universal constant
independent of microscopic couplings and persisting from very high to lowest temperatures over a
wide range of compounds \cite{Taillefer-planckian-2019} and doping \cite{Ramshaw-MottPlanckian-arxiv-2024}. It also shows "bad metallic" transport where perfect linear-in-temperature resistivity with the same slope persisting at highest experimentally
accessible temperature without saturation \cite{Taillefer-planckian-2019,Taillefer-annuphys-2019,Kivelson_1995_PRL_MIR}. 
By contrast, in conventional metals the electrical
resistivity saturates at high enough temperatures when scattering length of current-carrying
quasi-particles reaches the de-Broglie wavelength, so-called the Mott-Ioffe-Regel (MIR) limit \cite{Hussey_MIR_2004},
and the scattering rate is limited by quasi-particle interaction and disorder \cite{Mackenzie-Science,Zaanen-2015-QM-HTSC}. 
The microscopic origin of this exotic phase goes beyond the quasi-particle picture in Landau's Fermi liquid paradigm.

More recently, observations of 
frequency-to-temperature ($\hbar\omega/k_BT$) scaling from optical conductivity measurement \cite{George-NatComm-SM} and field-to-temperature ($B/T$) scaling in magnetoresistance extending over an extended doping range 
in various hole-doped \cite{hussey-incoherent-nature-2021} and electron-doped \cite{green-e-doped-cuprate-scale-invariant,Greene-nature-e-doped-cuprate-spin-fluc} cuprates strongly suggest that the Planckian metal is a “quantum critical phase” \cite{Hussey-Sci-Strangethanmetal}. 
Understanding the Planckian metal state is critical 
to understanding the mechanism for high-$T_c$ superconductivity in cuprates since the coefficient of the $T$-linear resistivity component tracks the doping dependence of  zero-temperature superfluid density \cite{hussey-incoherent-nature-2021} and therefore non-quasi-particles in this state play a crucial role in Cooper-pair condensation \cite{Hussey-Sci-Strangethanmetal}.

From theoretical perspectives, analytically solvable models in physics are very rare. 
The universal strange metal with Planckian-like behavior was shown to appear 
in the solvable SYK-like models in the large-$N$ limit where spatially random interacting electrons couple to a Fermi surface as well as in a large-$N$ approach of critical bosons coupled to conduction band with externally introduced spatially random interactions \cite{Patel-PRL-SM, Patel-2023-SYK-Sci}. 
Nevertheless, experimental evidence on correlation between impurity-induced disorder/randomness 
and the observed strange metal behavior in cuprates is yet to be addressed. Meanwhile, due to somewhat un-reallistic assumption of all-to-all spatially random interactions, the SYK-like models are distinct from the conventional Hubbard or $t$-$J$ models with non-disordered local interactions, known as the minimum effective models for describing cuprate phase diagram. 
It is therefore not clear whether the SYK-like models are applicable for describing 
cuprates. 
On the other hand, recent numerical study by quantum Monte Carlo \cite{deveraux-SM-Hubbard-Sci} 
and dynamical cluster approximation \cite{Tremblay-PNAS-2022} methods revealed the strange metal state with $T$-linear resistivity to appear in the conventional Hubbard model. 
Therefore, constructing an analytically solvable model within the conventional Hubbard or $t$-$J$ model becomes a more appealing and realistic alternative route for this exotic phase \cite{Arovas_2022_Review_Hubbard,Devereaux_science_2019_hubbard,Ogata_2008_ROPP,Punk-SBtJ-PRB}. 

In contrast to previous approaches 
\cite{Patel-PRL-SM,Patel-2023-SYK-Sci,Hartnoll-2015-NatPhys-diff,deveraux-SM-Hubbard-Sci,Tremblay-PNAS-2022,Karch-unparticles-2016,Nagaosa-1990-PRL}, 
here, we take an alternative and more realistic 
approach 
to the universal Planckian metal phase in the context of competing orders and quantum criticality in cuprate phase diagram: We develop 
a minimum solvable model 
within a large-$N$ multi-channel Kondo lattice model, derived from the heavy-fermion formulated 
conventional $t$-$J$ model \cite{YYC-ROPP-2025}. 
We show that this model is analytically solvable in the large-$N$ multi-channel limit. The ground state supports an extended universal quantum critical Planckian metal phase, made of (intrinsic) disordered local bosonic charge fluctuations coupled to both spinon and a heavy hole-like conduction-electron Fermi surfaces near a charge Kondo breakdown local quantum critical point, 
associated with the pseudogap-to-Fermi liquid transition within the effective $t$-$J$ model framework \cite{YYC-ROPP-2025}. 
Due to the local nature of electron scattering near criticality, this state shows the bad metallic transport property 
\cite{Kivelson_1995_PRL_MIR,Hussey_MIR_2004}. 
Our results are in excellent agreement with experimental observations \cite{George-NatComm-SM,zxshen-sicience-incoherent-cuprate,hussey-incoherent-nature-2021,michon-nature-2019-QC-cuprate}. 
Closely related approaches have been applied by some of us to successfully describe strange metal states appearing in heavy-fermion metals and superconductors close to the Kondo breakdown quantum critical point \cite{sm-phase-PNAS,Chang-2018-SM,YYC-SSC-2019,YYC-SM-115-NatComm}.

\noindent
\textbf{Large-$N$ multi-channel Kondo lattice model derived from conventional $t$-$J$ model$-$}
Our recent study in Ref.~\cite{YYC-ROPP-2025} based on perturbative renormalization group analysis demonstrates the existence of the Planckian metal phase in the heavy-fermion formulated slave-boson (conventional) $t$-$J$ model, which serves as our starting point here. 
Therein, the hopping $t$-term is mapped onto a Kondo-like coupling where 
disordered slave boson (holon $b$-field) interacts with fermionic spinon ($f$-field) band and the effective conduction electron band made of spinon-holon bound fermion 
($\xi$-field with $\xi_{ij}^{\sigma} = \sum_{\left<i,j\right>} b_i f_j^{\sigma}$) living 
on the bonds connecting nearest-neighbor sites $\left<i,j\right>$ \cite{Punk-PNAS-dimer,Punk-SBtJ-PRB}. 
The charge Kondo hybridization is dictated by the condensation of slave-boson.  
The anti-ferromagnetic Heisenberg exchange coupling $J$ term is decomposed into Resonating-Valence-Bond (RVB) spin liquid singlets in both particle-hole $\left( \chi_{ij} = \chi =\left<f_{i\sigma}^{\dagger} f_{j\sigma}\right> \right)$ and particle-particle $\left( d-\text{wave Cooper pairing } \Delta_{ij} = \Delta = \left< f_{i\uparrow}^{\dagger} f_{j\downarrow}^{\dagger} - f_{j\uparrow}^{\dagger} f_{i\downarrow}^{\dagger} \right> \right)$ channels. 
The Planckian metal phase appears as a result of 
critical local charge Kondo fluctuations (critical charge hopping) where disordered local bosons couple 
to a fermionic
spinon band and a heavy-electron band near a localized-deloclaized quantum critical point due to the competition
between the pseudogap phase (dominated by the RVB spin liquid via Heisenberg $J$ term) and FL phase (dominated by coherent charge hopping $t$-term) within our effective $t$-$J$ model. 
This approach is motivated by the striking similarity in strange metal phenomenology between cuprates \cite{Taillefer-annuphys-2019,hussey-incoherent-nature-2021} 
and heavy-fermion Kondo lattice systems \cite{Friedemann-PNAS-2010}, 
in which the Fermi surface volume reconstructs over the entire strange metal region in both systems. It 
well captures the relations among the Planckian metal state with quantum critical scaling in transport and the pseudogap, $d$-wave superconducting, and Fermi liquid phases of cuprates as observed.  
 
Here, we develop a minimum solvable model for Planckian metal 
based on 
the model in Ref.~\cite{YYC-ROPP-2025} 
close to charge Kondo breakdown local criticality 
by focusing on the most important charge fluctuations in the hopping $t$-term (effective Kondo coupling) and neglecting the pairing ($\Delta$-field) fluctuations of RVB spin liquid. 
We generalize this minimum model to an analytically solvable large-$N$ multi-channel ($K$-channel) Kondo lattice model with $N$ flavors of spin and $K$ screening channels with following action \cite{YYC-ROPP-2025} 
     \begin{align}
         S =& -\sum_{k\sigma} f_{k\sigma}^{\dagger} \mathcal{G}_{f,\sigma}^{-1}(k) f_{k\sigma} - \sum_{ka} b_{ka}^{\dagger} \mathcal{G}_{b,a}^{-1}(k) b_{ka} \nn  
         & - \sum_{k\sigma a} \xi_{k\sigma a}^{\dagger} \mathcal{G}^{-1}_{\xi,\sigma a}(k) \xi_{k\sigma a} \nn
         & +\frac{g}{\sqrt{\beta N_{s}N}}\sum_{kp\sigma a}\left(f_{k\sigma}^{\dagger}b_{pa}^{\dagger}\xi_{k+p,\sigma a}+H.c.\right) \;,  
    \label{eq:action-S}
     \end{align}
where $\mathcal{G}_{f,\sigma} (k)= (ik_n-\varepsilon_{\bm{k}}-\mu_f)^{-1}$ and $\mathcal{G}_{b,a}(k) = (ik_n - \lambda)^{-1}$ are initial (bare) Green's functions of the $f$-spinon and the slave-boson $b$ field, respectively;  
$\mathcal{G}_{\xi,\sigma a} (k) = 
\left[ \zeta (ik_n + \varepsilon_{\bm{k}} - \mu_{\xi}) - i\Gamma_0 \sgn(k_n) \right]^{-1}$ 
(with $\zeta=g^2\rho_0/D=g^2/D^2$, $\Gamma_0=\pi D$, $D \sim J$ being $f$-spinon bandwidth)  
is the initial Green's function of the spinon-holon bound $\xi$-fermion.  
Here $\varepsilon_{\bm{k}}=-2\chi\left(\cos k_{x}+\cos k_{y}\right)$, 
$\mu_f\,(\mu_{\xi})$ is the effective chemical potential for 1/2-filled $f$-spinon ($\delta$-hole-doped $\xi$-fermion), respectively; $\lambda$ is the Lagrange multiplier to fix the local single-occupation constraint: $f_{\sigma i}^{\dagger} f_{\sigma i} + b_i^{\dagger} b_i = 1$; and the effective Kondo coupling $g$ is related to the hopping term $t$ by $g \sim t\sqrt{\delta}$. 
In the strange metal region the slave bosons are not condensed, and therefore act as local disordered impurity scatterer. 
Since $\xi$-fields carry both spin and positive charge, they play the role of hole-like conduction electrons, conducting electrical transport. 
The dynamics and dispersion of $\xi$-field beyond static dispersionless mean-field level are generated by 2$^{\text{nd}}$-order perturbation in hopping term, through which a dissipative constant decay rate $\Gamma_0$ in $\mathcal{G}_{\xi,\sigma a} (k)$ is also generated due to the local impurity-like scattering among $\xi$, $f$-spinon and slave boson (Supplementary Materials (SM) \ref{Methods} and \ref{DD-RPA}). 
Note that disorder is crucial for realizing the Planckian metal. The non-condensed slave boson here plays the role of "intrinsic" disorder in the effective Kondo term though hopping parameter $t$ itself is non-random, in contrast to the "external" disorder introduced in the SYK models \cite{Patel-PRL-SM, Patel-2023-SYK-Sci}. 
In Eq.~(\ref{eq:action-S}), we generalize the symmetry group of the spin sector from SU(2) to SU(N) $\times$ SU(K) 
accompanying with the change in the spin index $\sigma= \pm 1,\cdots,\pm\frac{N}{2}$ and the Kondo screening channel from 1 to $K$ with $a = 1,\cdots,K$. In the large-$N$ limit, 
we send $N,\;K\to\infty$ but keep $\kappa \equiv K/N$ finite. (Here, we set $\kappa = 1/2$, which recovers physical $s=1/2$ SU(2) limit with $N=2$ and $K=1$.) 
Our model in Eq.~(\ref{eq:action-S}) shows the U(1) gauge symmetry: $f_{i\sigma} \to f_{i\sigma} e^{i\theta_i},\,b_i \to b_i e^{i\theta_i},\,\chi_{ij} \to \chi_{ij} e^{i(\theta_i-\theta_j)}$ and $\xi_{ij,\sigma} \to \xi_{ij,\sigma} e^{i(\theta_i+\theta_j)}$.

The one loop RG analysis on Eq.~(\ref{eq:action-S}) is carried out in the large-$N$ multi-channel limit, leading to the scaling equation for the effective Kondo coupling $g \rho_0$ with $\rho_0 = 1/D$ being the conduction electron density of states (SM \ref{app:RG})
\begin{align}
    \frac{d(g\rho_0)}{d\ell} = -\left(\frac{\epsilon}{2}\right) \left(g\rho_0\right) + \kappa \left(g\rho_0\right)^3\;. 
    \label{eq:g-beta}
\end{align}

By Eq.~(\ref{eq:g-beta}), the Kondo breakdown local quantum critical point occurs at $g_c \rho_0= \sqrt{\epsilon/(2\kappa)}$, dependent of the value of $\kappa$. The hallmarks of this local quantum critical point are: (i) the Fermi surface reconstruction: the Fermi surface volume changes from a smaller value ($p$) for $g < g_c$ to a larger value ($1 + p$) for $g > g_c$ due to condensation of slave-boson $b$-field, and (ii) the momentum-insenstive charge susceptibility. 
The former is consistent with the change observed in Hall coefficient across the strange metal \cite{hussey-incoherent-nature-2021,Taillefer-annuphys-2019}, while the latter supportive by the momentum-independent continuum observed in M-EELS \cite{Mitrano2018,Abbamonte_PRX_2019}. 
We will show below that for $g< g_c$ where slave-boson is non-condensed, Eq.~\eqref{eq:action-S} is exactly solvable in the large-$N$ multi-channel limit, and the solution supports a universal quantum critical Planckian metal phase.

\begin{figure}	
    \centering
    \includegraphics[width=0.7\textwidth]{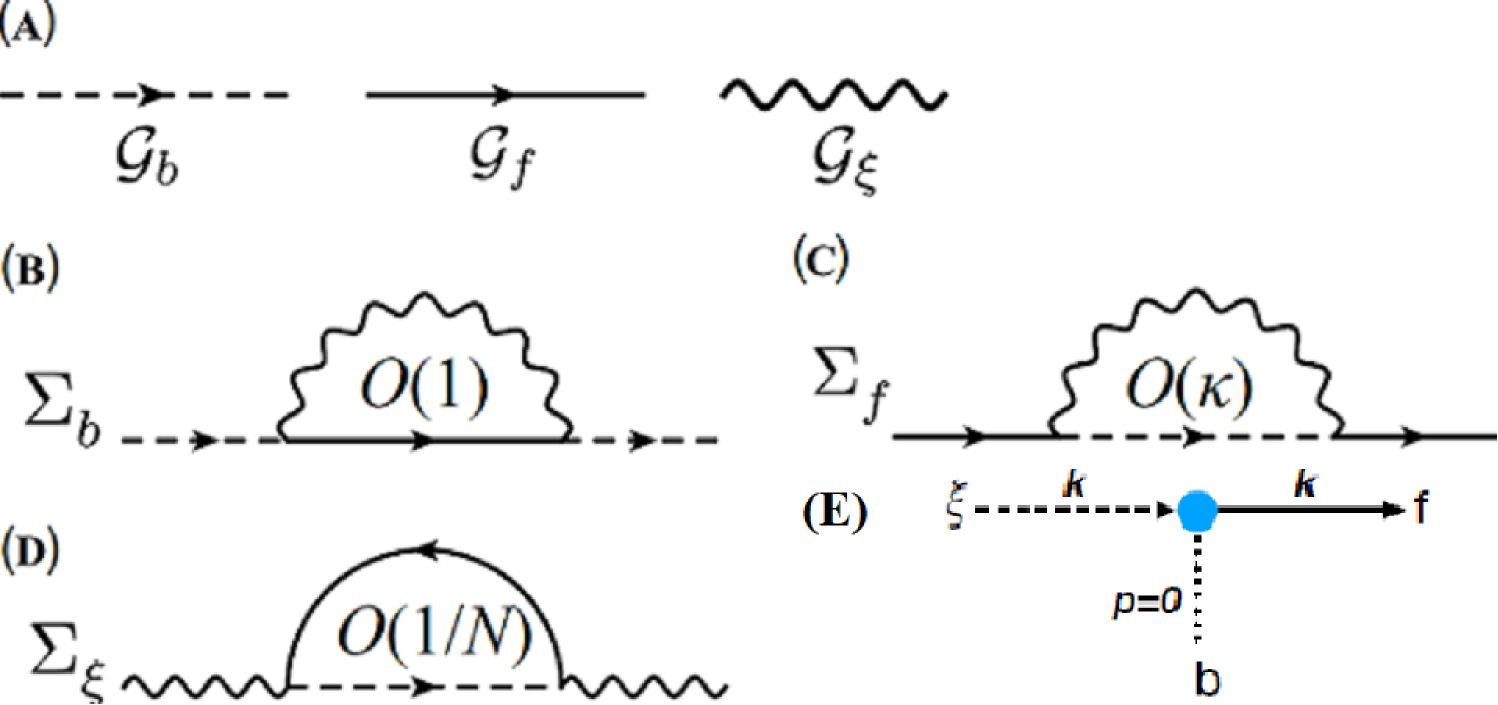}
    \caption{\textbf{The propagators and Feynman diagrams.} (A) The graphical representation of the propagator of various fields/operators. (B)-(D) Feynman diagrams of the self-energy contributing from the perturbation of $H_t$ up to the one-loop order. (E) Scattering process within our model.}
    \label{fig:feyn-diag}
\end{figure}

\noindent
\textbf{Exact solution of Planckian metal phase in large-$N$ limit$-$}
In the absence of boson condensation $\langle b \rangle =0$ (i.e., for $g<g_c$ in Eq. \eqref{eq:g-beta}), the Dyson equations for the local Green functions $G_{\xi,\sigma a}$, $G_{f,\sigma}$, $G_{b,a}$ and the local self-energies of a certain spin $\sigma$ and channel $a$ read
\begin{align}
    \left[G_{\xi,\sigma a}(\omega)\right]^{-1} &=  \left[\mathcal{G}_{\xi,\sigma a}(\omega)\right]^{-1} - \Sigma_{\xi,\sigma a} (\omega) \; ,  \nn
    \left[G_{b,a}(\omega)\right]^{-1} & =  \left[\mathcal{G}_{b, a} (\omega)\right]^{-1}- \Sigma_{b,a}(\omega) \; , \nn
    \left[G_{f,\sigma}(\omega)\right]^{-1} & =  \left[\mathcal{G}_{f,\sigma}(\omega) \right]^{-1}- \Sigma_{f,\sigma}(\omega)\; ,  
\end{align}
and
\begin{align}
    \Sigma_{\xi,\sigma a} (\omega) & = - \frac{1}{N} \sum_{\nu} G_{f,\sigma} (\nu) G_{b,a} (\omega-\nu) \;, \nn
    \Sigma_{b,a} (\omega)  &= 2{g}^{2} \sum_{\nu,\sigma}  G_{f,\sigma}(\nu) G_{\xi,\sigma a} (\omega+\nu) \;, \nn
    \Sigma_{f,\sigma} (\omega)  &=  -\kappa g^2 \sum_{\nu,a} G_{\xi,\sigma a} (\omega+\nu) G_{b,a} (\nu)\; ,
\end{align}
where $G_{\xi,f,b}(\omega) = \sum_k G_{\xi,f,b}(\omega,k)$ and 
$\Sigma_{f,\xi,b}(\omega) = \sum_k \Sigma_{f,\xi,b}(\omega,k)$ 
are the local Green functions and the local self-energies, respectively. 
The self-energies shown above are pictorially demonstrated in Figs. \ref{fig:feyn-diag}(B)-\ref{fig:feyn-diag}(D). 
Since $\Sigma_{\xi,\sigma a} \sim O(1/N)$, 
the dressed Green's function $G_{\xi}$ reduces to the initial Green's function 
$\mathcal{G}_{\xi}$ in the large-$N$ limit, $\lim_{N \rightarrow \infty} G_{\xi} = \mathcal{G}_{\xi}$. 
Due to local character in nature, we find the self-energy correction to the bare slave-boson Green function is negligible in the wide-band limit, $\Sigma_b(\omega)\to 0$ (SM \ref{localbf}). 
Combining this result and our Ansatz—the complete local slave-boson field with $\Sigma_b=0$, we obtain $\Sigma_f(\omega)$ via Eq. (4), reads: 
$\Sigma_f(\omega) = \Sigma_f^{'}(\omega) + i\Sigma_f^{''}(\omega)$ where 
$\Sigma_f^{''}(\omega) = -\alpha\sgn(\omega) +\bar{\zeta}|\omega|$ with $\alpha \approx 2\kappa D/3$ and $\bar{\zeta} = 2\kappa/\pi$, and $\Sigma_f^{'}(\omega)$ obtained from $\Sigma_f^{''}(\omega)$ via Kramers-Kronig relation. 
To check the consistency of our Ansatz $\Sigma_b = 0$, we substitute $\Sigma_f$ into $G_f$, followed by substituting $G_f$ and $G_{\xi}$ into Eq. (4) for $\Sigma_b$. 
We find indeed that $\Sigma_b$ vanishes; therefore, in the large-$N$ limit the slave boson is fully localized with a dispersionless energy level, reminiscent of a local impurity (SM \ref{localbf}). 
The self-consistency 
is reached by plugging $\Sigma_f$, $\Sigma_b$, and $\Sigma_{\xi}$ above into Eq. (3) and Eq. (4) (SM \ref{SBtJfor}). 

\begin{figure}	
    \centering
    \includegraphics[width=0.7\textwidth]{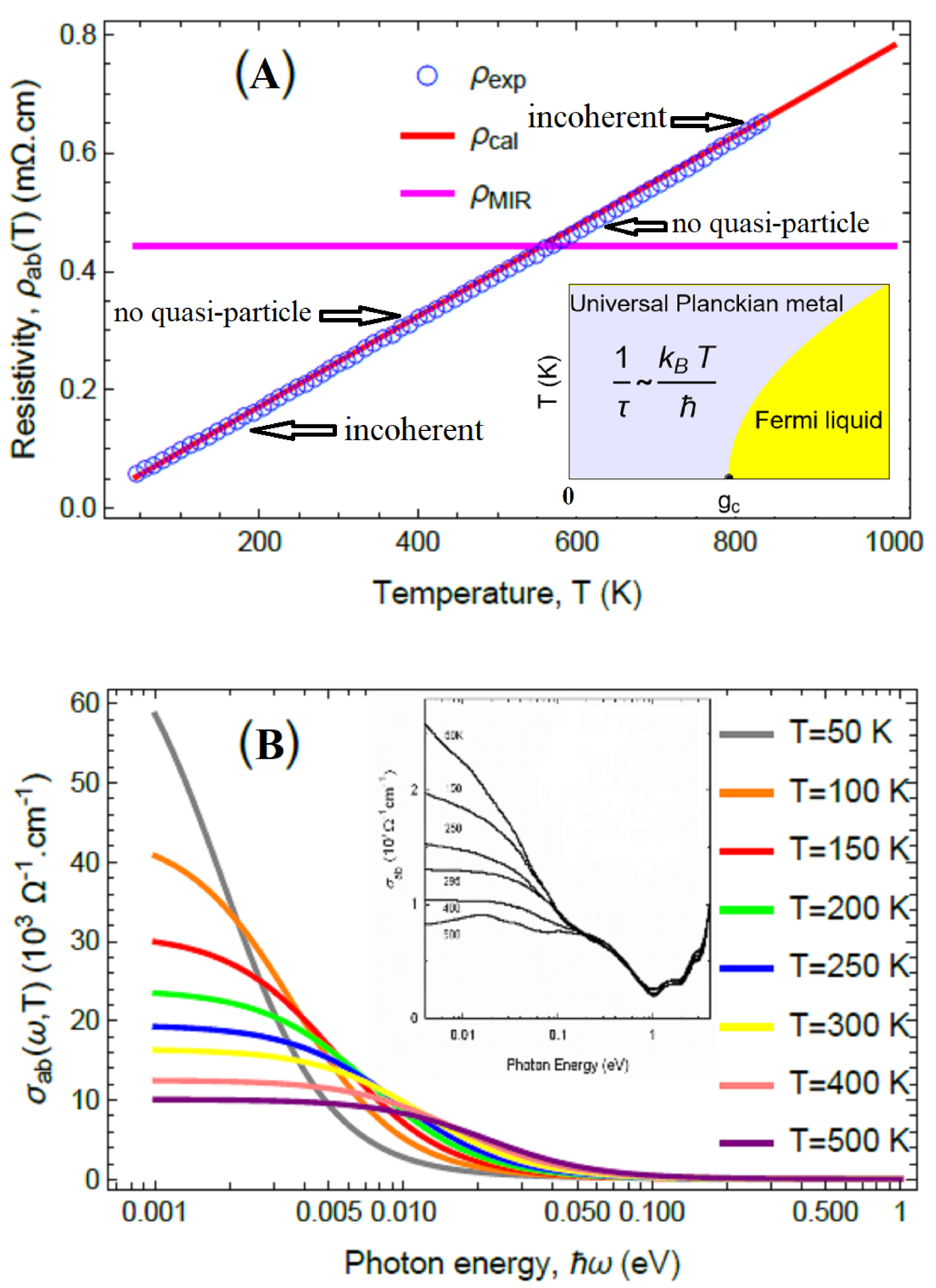}
    \caption{\textbf{In-plane dc-resistivity and ac-conductivity of LSCO.} (A) In-plane dc-resistivity is calculated from the scattering rate predicted within our theory (Eq. (\ref{eq:uqc-scaling})) at $\delta=0.15$ (SM \ref{dc-ac}). The calculated resistivity (red line) is compared with the experimental data (blue circles) reproduced from \cite{Takagi_1992_LSCO_PRL}. The horizontal magenta line shows $\rho_{\rm MIR}$ for the normal metal. The inset figure displays the seperation between the universal Planckian-metal and Fermi-liquid regimes in the ($T,g$) phase diagram. (B) In-plane ac-conductivity as a function of photon energy at different temperatures is computed at $\delta=0.24$ using the scattering rate given by Eq. (\ref{eq:uqc-scaling}) (SM \ref{dc-ac}). The inset figure is the experimental data for La$_{1.9}$Sr$_{0.1}$CuO$_4$ taken from Ref.~\cite{Hussey_MIR_2004} (by courtesy of Nigel Hussey).}
    \label{fig:DC-AC}
\end{figure}

We now apply the exact solution of our large-$N$ model Eq.~\eqref{eq:action-S} for electron transport. The gauge-invariant physical electron operator in the Planckian phase is related to the spinon-holon bound fermion by: 
$c_i^{\sigma} = \sqrt{2}\sum_{<ij>} \xi_{ij}^{\alpha} b_i^{\dagger} b_j^{\dagger}$. The electron scattering (relaxation) rate $1/\tau_{\rm re}$ is given by the imaginary part of the $T$-matrix (T=0): $1/\tau_{re}=1/\tau_c\equiv-NT^{\prime\prime}(\omega)$ with $T^{\prime\prime}(\omega)= -2\Sigma_c^{\prime\prime}(\omega)$, and $\Sigma_c^{\prime\prime}=\frac{1}{2}\Sigma_{\xi}^{\prime\prime}=-\frac{A}{2N}-\frac{2\kappa\varsigma}{N} |\omega|$ (A is a constant, SM \ref{SBtJfor} and Eq. \eqref{eq:Im-Sigma-xi-final}) the c-electron self-energy featuring universal Marginal Fermi Liquid behavior \cite{YYC-ROPP-2025}.  
Since c-electron self-energy is completely local, the transport time $\tau_{\rm tr}$ is given by the relaxation time $\tau_{\rm re}$, $\tau_{\rm tr} \sim \tau_{\rm re}$ \cite{George-NatComm-SM}. The scattering rate at a finite frequency and temperature in the quantum critical scaling regime shows the universal quantum critical scaling \cite{YYC-ROPP-2025} 
\begin{align}
    \frac{\hbar/\tau -\hbar/\tau_0}{k_B T} = \frac{4\kappa}{\pi} x \coth \frac{x}{2},
    \label{eq:uqc-scaling}
\end{align}
where $x = \hbar\omega/k_BT$. 
The c-electron self-energy
features universal Marginal Fermi Liquid behavior \cite{YYC-ROPP-2025}, in which charge transport is incoherent as the width of the peak in single-particle spectral function, proportional to $\text{max}(\omega,T)$, is much broader than the width of a Lorentzian peak $\text{max}(\omega^2,T^2)$ of a Fermi liquid. This incoherent transport persists at all frequencies and temperatures as long as the system is in the quantum critical scaling regime (Fig.~\ref{fig:DC-AC} (A)). 
By Eq. (5), in the AC-limit $\omega \gg T$ scattering rate is linear-in-$\omega$, $1/\tau \sim (4\kappa/\pi) |\omega|$, while in the DC-limit $\omega \ll T$, universal Planckian scattering rate is reached, $1/\tau_P = \alpha_P k_B T/\hbar$ with  $\alpha_P = 8\kappa/\pi = 4/\pi \approx 1.27$. 
The perfect Planckian $T$-linear resistivity over the scaling regime is reached via Drude formula: $\rho_P = (ne^2/m^*)\,1/\tau_P$ (Fig.~\ref{fig:DC-AC} (A) the fit to $T$-linear resistivity of LSCO in Ref.~\cite{Takagi_1992_LSCO_PRL}). 
Due to a cancellation of coupling constant $g$ in $\Sigma_{f/\xi}$, the universal Planckian scattering rate in Eq. (5) is independent of coupling constant and persists over an extended parameter range for $g<g_c$ (Fig.~\ref{fig:DC-AC} (A) Inset), in excellent agreement with various observations in transport of cuprates over an extended overdoped regime \cite{Taillefer-planckian-2019, Ramshaw-MottPlanckian-arxiv-2024}. 
In the quantum critical regime thermal energy $k_BT$ is the only relevant (largest) energy scale; it therefore naturally gives rise to the Planckian bound for scattering rate $1/\tau_P \sim k_BT/\hbar$ as observed in cuprates \cite{Zaanen-2004-nature}. Moreover, we offer here an inelastic scattering mechanism for momentum and energy relaxation even in the absence of Umklapp scattering 
(often required to reach thermal equilibrium for scatterings in conventional metals): The electrical current carried by $\xi$-field is relaxed by decaying into local impurity-like charge represented by non-condensed slave-boson and the charge-neutral $f$-spinon current via these local fluctuating critical charge-Kondo coupling, contributed to the local self-energy. 
The same coupling also gives rise to energy relaxation. 
This local quantum criticality scenario for the universal Planckian metal is distinct from previous ones.

\noindent
\textbf{The Planckian bad metal and the Mott-Ioffe-Regel limit$-$}
In ordinary metals, the dc-electrical resistivity is bounded by the MIR limit, $\rho < \rho_{\rm MIR}$, 
when the mean free path $l$ at least equals the Fermi wavelength $\lambda_F$, $l \ge \lambda_F = 2\pi/k_F$ or $k_F l \ge 2\pi$ \cite{Kivelson_1995_PRL_MIR} (an estimated $\rho_{\rm MIR}$ for the normal metal (SM \ref{app:rMIR}) is shown in Fig.~\ref{fig:DC-AC} (A)). 
The bad metallic behavior in the Planckian metal is well accounted for within our theory for the following reasons. Firstly, due to the local nature of the critical charge Kondo fluctuations in our model, $l \to 0$, leading to $\rho_{\rm MIR}^{\rm P} \propto \frac{v_F}{l} \to \infty$ (SM \ref{app:rMIR}); 
therefore, the $\rho_{\rm MIR}^{\rm P}$ in this case cannot be reached in the experimentally accessible finite temperature range. 
Secondly, the Planckian state (Eq.~(5)) with incoherent transport without quasi-particles persists for all temperatures and frequencies as long as they are in the quantum critical scaling regime, which can in principle go beyond the MIR bound (Fig.~\ref{fig:DC-AC} (A)). 
Nevertheless, the signature of MIR bound still shows up in ac-conductivity as observed in overdoped cuprates \cite{Millis_1987_PRB,Hussey_MIR_2004} where 
$\text{Re}\left[ \sigma(\omega,T) \right] = \frac{ne^2}{m_b} 
{\tau(\omega,T)}/
\left[ {1 + \left( \frac{m^*(\omega,T)}{m_b} \right)^2 \omega^2 \tau^2(\omega,T)} \right]$ 
(SM \ref{dc-ac} and Fig.~\ref{fig:DC-AC} (B)) 
tends to saturate in high temperature and low frequency, followed by a shift of spectral weight to higher frequencies. The saturated value for ac-conductivity in the static limit approximately  corresponds to the MIR bound:  $\sigma (\omega\to 0, T) \sim 1/ \rho_{\rm MIR}$. Remarkably, these qualitative signatures of MIR bound are 
well captured within our model by computing the ac-conductivity $\sigma(\omega,T)$ via Eq. (5) (Fig.~\ref{fig:DC-AC} (B)).

\noindent
\textbf{Discussion and outlook$-$}
Here, without Umklapp scattering, we offer a distinct relaxation mechanism via local charge Kondo fluctuations where electrical charge current decays into charge-neutral spinon current and local charge fluctuations despite forward scattering as shown in Fig.~1 (E). 
The $f$-spinons in U(1) FL* spin-liquid phase are deconfined due to the existence of a spinon Fermi surface, and are thus stable against U(1) gauge field fluctuations \cite{SSLee-PRB-2008-stablility,Hermele-stability-spinliquid}. 
Though our calculations are done in the large-$N$ limit, we find that exact cancellation of the 
Kondo coupling in electron self-energy, leading to the Planckian scattering rate, survives at any finite $N$. The large-$N$ approach here mainly serves as a systematic way to select a consistent set of diagrams to be resummed based on the saddle point of our effective action, which makes the theory analytically solvable. 
The contributions from U(1) gauge field fluctuations to our large-$N$ solutions are negligible in the large-$N$ limit \cite{Senthil-prl-fractionalized-FL,SSLee-PRB-2008-stablility}. At finite-$N$, due to the much heavier $\xi$-band compared to that of $f$-spinon, we find that their contributions are subleading and are negligible \cite{YYC-ROPP-2025}.

Our mechanism shows a broader implication for the strange metal state observed in other correlated unconventional superconductors, such as: nickelate superconductors \cite{Harold-nickelate-QC-SM} where a T-linear resistivity was observed in the quantum critical-fan-like region above the superconducting phase, similar to that in cuprates. The Kondo-Hubbard model was proposed to serve as a promising minimum microscopic model for this material \cite{Devereaux-nickelate-kondo-hubbard}, where the local Kondo-breakdown transition, similar to our case, may occur. The combined quantum critical bosonic charge fluctuations in the correlated electron hopping and Kondo interaction near local quantum criticality may provide a mechanism for its strange metal behavior.

%


%

\section*{Acknowledgments} 
C.H.C. acknowledges discussions with P.H. Chou. 
This work is supported by the National Science and Technology Council (NSTC) Grants 110-2112-M-A49-018-MY3, the National Center for Theoretical Sciences of Taiwan, Republic of China (C.H.C.). 
Y.Y.C. acknowledges the financial support from The 2023 Postdoctoral Scholar 13 Program of Academia Sinica, Taiwan. 
C.H.C. acknowledges the hospitality of Aspen Center for Physics, USA and Kavli Institute for Theoretical Physics, UCSB, USA where part of the work was done. 
This research was supported in part by grant NSF PHY-2309135 to the Kavli Institute for Theoretical Physics (KITP).

\clearpage 

\newpage

\renewcommand\thesection{S\Roman{section}}
\renewcommand{\thefigure}{S\arabic{figure}}
\renewcommand{\thetable}{S\arabic{table}}
\renewcommand{\theequation}{S\arabic{equation}}
\renewcommand{\thepage}{S\arabic{page}}
\setcounter{figure}{0}
\setcounter{table}{0}
\setcounter{equation}{0}
\setcounter{page}{1} 


\begin{center}
\section*{Supplementary Materials for\\ \scititle}

Yung-Yeh Chang,
Khoe Van Nguyen,
Kimberly Remund,
Chung-Hou Chung$^{\ast}$\\ 
\small$^\ast$Corresponding author. Email: chung0523@nycu.edu.tw\\
\end{center}

\subsubsection*{This PDF file includes:}
Methods: The large-$N$ multi-channel heavy-fermion formulated slave-boson $t$-$J$ (HFSBtJ) model\\
Dynamics and/or dispersion of $\xi$-fermion and slave-boson $b$-field\\
Renormalization group (RG) analysis: scaling dimension, self-energy corrections, vertex corrections, the renormalization $z$-factors and the RG $\beta$ functions, the RG scaling equations ($\beta$ functions)\\
Demonstration of $\Sigma_b(\omega) \to 0$\\
The large-$N$ multi-channel HFSBtJ formulation: scaling of AC resistivity\\
Estimation of $\rho_{\text{MIR}}$\\
Calculations of in-plane dc-resistivity and ac-conductivity of LSCO\\
Figure S1



\newpage


\section{Methods: The large-$N$ multi-channel heavy-fermion formulated slave-boson $t$-$J$ (HFSBtJ) model}\label{Methods}

Our starting point is the heavy-fermion formulated slave-boson $t$-$J$ (HFSBtJ) Hamiltonian on 2D lattice which we developed recently
\begin{align}\label{eq:SBtJH}
    H_{t} &=-t\sum_{\langle i,j \rangle , \sigma} c_{i\sigma}^{\dagger} c_{j\sigma}-\mu \sum_{i\sigma} c_{i\sigma}^{\dagger} c_{i\sigma}, \nn
     H_J  & = J_{H} \sum_{\langle i,j\rangle} \bm{S_{i}} \cdot\bm{S_{j}}\;,
\end{align}
where $(t,\, \mu, \, J_H)$ denotes the (hopping strength, chemical potential, Heisenberg coupling). 
Under the slave-boson representation $c_{i\sigma}^\dagger \to  f^\dagger_{i\sigma} b_i$ with $f_{i\sigma} \, (b_i)$ being fermionic charged-neutral spinon (bosonic spinless charged holon), we further factorize the $H_t$ and $H_J$ terms via Hubbard-Stratonovich transformation following Ref. \cite{Punk-SBtJ-PRB},
\begin{align}
    H_{t} &\to \frac{t}{\sqrt{N}}\sum_{\langle i,j\rangle,\sigma} \left[\left(f_{i\sigma}^{\dagger}b_{j}^{\dagger}+f_{j\sigma}^{\dagger}b_{i}^{\dagger}\right)\xi_{ij,\sigma}+H.c.\right] \;.
\end{align}
The action of the model is then described by
 \begin{align}
         S =& -\sum_{k\sigma}f_{k\sigma}^{\dagger}\left[\mathcal{G}^{\rm{B}}_{f,\sigma} (k)\right]^{-1}f_{k\sigma}-\sum_{ka}b_{ka}^{\dagger}\left[\mathcal{G}^{\rm{B}}_{b,a}(k)\right]^{-1}  b_{ka} \nn  
         & -\sum_{k\sigma a} \xi_{k\sigma a}^{\dagger} \left[\mathcal{G}^{\rm{B}}_{\xi,\sigma a}(k)\right]^{-1} \xi_{k\sigma a} \nn
         & +\frac{g}{\sqrt{\beta N_{s}N}}\sum_{kp\sigma a}\left(f_{k\sigma}^{\dagger}b_{pa}^{\dagger}\xi_{k+p,\sigma a}+H.c.\right), 
    \label{eq:action-S-2}
     \end{align}
     with  $\mathcal{G}^{\rm{B}}_{f,\sigma} (k)= (ik_n-\varepsilon_{\bm{k}})^{-1}$, $\mathcal{G}^{\rm{B}}_{\xi,\sigma a} (k) = -\frac{2}{g}$, $\mathcal{G}^{\rm{B}}_{b,a}(k) = (ik_n - \lambda)^{-1}$  denoted the ``\textit{bare}" Green's functions at the mean-field level (thus labeled with superscript ``$\rm{B}$")  for certain spin $\sigma$ and channel $a$ species, distinguish from the RPA-corrected initial Green functions as defined in Eq.~\eqref{eq:action-S}. However, in its bare Green's functions $\mathcal{G}^{\rm B }$ of Eq.~\eqref{eq:action-S-2}, the $\xi$ field does not reflect its intrinsic itinerant nature as it lacks the explicit dynamical properties and energy dispersion. To give the $\xi$ field the itinerant electron nature, we employ the RPA to generate its dynamics and dispersion, resulting in the action given in Eq.~\eqref{eq:action-S} (Ref.~\cite{YYC-ROPP-2025} and Eqs.~(S17) and (S18) in Supplementary Materials \ref{DD-RPA}).








\section{Dynamics and/or dispersion of $\xi$-fermion and slave-boson $b$-field}\label{DD-RPA}

Below we generate dynamics and dispersion of $\xi$-band by Random Phase Approximation (RPA) approach \cite{YYC-ROPP-2025}. 
At the bare level, the $\xi$ fermion does not have dynamics. 
The dynamics and dispersion of the $\xi$-fermion are generated perturbatively by calculating the leading non-trivial self-energy correction $\Sigma_{\xi}$ via RPA approach where high energy modes in $\Sigma_{\xi}$ are integrated out \cite{Chang-2018-SM,YYC-SSC-2019}. 
We first evaluate $\Sigma_{\xi}$: The Green's function of the $\xi$ fermion to the second order in hopping term is given by
\begin{align}
    \mathcal{G}_{\xi}^{(2)}(\bm{k},\omega_{n})	& =-\frac{g^{2}}{\beta}\sum_{p,q,\sigma}\sum_{p^{\prime},q^{\prime},\sigma^{\prime}}\langle\xi_{\bm{k}\sigma}(\omega_{n})\xi_{p+q,\sigma}^{\dagger}b_{q}f_{p\sigma}f_{p^{\prime}\sigma^{\prime}}^{\dagger}b_{q^{\prime}}^{\dagger}\xi_{p^{\prime}+q^{\prime},\sigma^{\prime}}\xi_{\bm{k}\sigma}^{\dagger}(\omega_{n})\rangle_{0} \nn
	& =-\frac{g^{2}}{\beta}\sum_{p,q,\sigma}\sum_{p^{\prime},q^{\prime},\sigma^{\prime}}\left\langle \xi_{\bm{k}\sigma}(\omega_{n})\xi_{p+q,\sigma}^{\dagger}f_{p\sigma}f_{p^{\prime}\sigma^{\prime}}^{\dagger}\xi_{p^{\prime}+q^{\prime},\sigma^{\prime}}\xi_{\bm{k}\sigma}^{\dagger}(\omega_{n})\right\rangle _{0}\left\langle b_{q}b_{q^{\prime}}^{\dagger}\right\rangle _{0} \nn
	& =-\frac{g^{2}}{\beta}\sum_{p} \left( \mathcal{G}_{\xi}^B(\bm{k},\omega_{n}) \right)^{2} \mathcal{G}_{f}(\bm{p},p_{n})\mathcal{G}_{b}(\bm{k-p},\omega_{n}-q_{n}).
\end{align}
Thus, via the Dyson equation, $G_{\xi}^{-1}=\left(\mathcal{G}_{\xi}^B\right)^{-1}-\Sigma_{\xi}$, the full $G_{\xi}$ up to the second order approximation, $  G_{\xi}(\bm{k},\omega_{n})	 \approx\mathcal{G}_{\xi}^B(\bm{k},\omega_{n})+\mathcal{G}_{\xi}^{(2)}(\bm{k},\omega_{n}) $, leads to
	\begin{align}
	    G_{\xi}^{-1}(\bm{k},\omega_{n})=\left(\mathcal{G}_{\xi}^B(\bm{k},\omega_{n})\right)^{-1} - \Sigma_{\xi}(\bm{k},i\omega_{n}),
	\end{align}
    where we obtain 
    \begin{equation}
        \Sigma_{\xi}(\bm{k},\omega_{n})=-\frac{g^{2}}{\beta}\sum_{\bm{p},p_{n}}\mathcal{G}_{f}(\bm{p},p_{n})\mathcal{G}_{b}(\bm{k}-\bm{p},\omega_{n}-p_{n}).
    \end{equation}
To acquire the dispersion of the $\xi$ fermion, we only focus on real part of $\Sigma_{\xi}$. We integrate out the higher energy modes of the conduction electron, leading to 
\begin{align}
    \Sigma_{\xi}(\bm{k},\omega_{n}) = & -\frac{g^{2}}{\beta}\sum_{\bm{p},p_{n}}\mathcal{G}_{f}(\bm{p},p_{n})\mathcal{G}_{b}(\bm{k}-\bm{p},\omega_{n}-p_{n}) \nn
	= & \sum_{\bm{p}}\left(\frac{g^{2}}{i\omega_{n}-\lambda-\varepsilon_{\bm{p}}}\right)+\sum_{\bm{p}}^{\sim}\left(\frac{g^{2}}{-i\omega_{n}+\varepsilon_{\bm{p}}+\lambda}\right),
\end{align}
where, in the above equation, $\tilde{\sum}_{\bm{p}}$ means summing over the relatively higher energy (quasi-momenta) states whose energy levels are slightly below the Fermi surface.

Letting $i\omega_{n}\to\omega+i\epsilon$ with $\epsilon>0$ being an infinitesimal number. The real part of first term is
\begin{align}
    	\sum_{\bm{p}}\left(\frac{g^{2}}{\omega-\lambda-\varepsilon_{\bm{p}}}\right) 
	= & -g^{2}\rho_{0}\left(\int_{-D}^{-D+\varepsilon_{\bm{k}}}+\int_{D-\varepsilon_{\bm{k}}}^{D}\right)\frac{d\varepsilon}{-\omega+\lambda+\varepsilon} \nn
	 \approx & -g^{2}\rho_{0} \left[\ln\left(1+\frac{\omega-\lambda-\varepsilon_{\bm{k}}}{D}\right)-\ln\left(1+\frac{-\omega+\lambda-\varepsilon_{\bm{k}}}{D}\right)\right] \nn
	 \approx &  \frac{-2g^{2}\rho_{0}}{D}\left(\omega-\lambda\right),
\end{align}
and the real part of the second term is given by
\begin{align}
     \sum_{\bm{p}}^{\sim}\left(\frac{g^{2}}{-\omega+\varepsilon_{\bm{p}}+\lambda}\right) 	
    = & g^{2}\rho_{0}\int_{-D}^{-D+\varepsilon_{\bm{k}}}\frac{d\varepsilon}{-\omega+\varepsilon+\lambda} \nn
    =& g^{2}\rho_{0}\ln\left(\frac{-\omega+\lambda-D+\varepsilon_{\bm{k}}}{-\omega+\lambda-D}\right) \nn
	 \approx & \frac{g^{2}\rho_{0}}{D}\left(\omega-\lambda-\varepsilon_{\bm{k}}\right).
\end{align}
The real part of $\Sigma_{\xi}(\omega+i\epsilon,\bm{k})$  then reads
\begin{align}
    \Sigma_{\xi}(\omega+i\epsilon,\bm{k})\approx-\frac{g^{2}\rho_{0}}{D}\varepsilon_{\bm{k}}-\frac{g^{2}\rho_{0}}{D}\left(\omega-\lambda\right).
\end{align}

Next, we evaluate the imaginary part of $\Sigma_{\xi}$, denoted as $\Sigma_{\xi}^{\prime\prime}$. Using $(x\pm i\epsilon)^{-1}=\mathcal{P}(\frac{1}{x})\mp i\pi\delta(x)$ and assuming $\left|\omega-\lambda\right|\leq D$, we have 
\begin{align}
     \Sigma_{\xi}^{\prime\prime}	(\omega\pm i\epsilon,\bm{k})   
     = &  \sum_{\bm{p}}\left(\frac{g^{2}}{\omega-\lambda-\varepsilon_{\bm{p}}\pm i\epsilon}\right)^{\prime\prime}+\sum_{\bm{p}}^{\sim}\left(\frac{g^{2}}{-\omega+\varepsilon_{\bm{p}}+\lambda\pm i\epsilon}\right)^{\prime\prime} \nn
	= & \mp\pi g^{2}\rho_{0} \int_{-D}^{D}d\varepsilon\delta\left(\omega-\lambda-\varepsilon\right)\pm\pi g^{2}\rho_{0}\int_{-D}^{0}\delta\left(\lambda+\varepsilon-\omega\right) \nn
	= & \begin{cases}
\mp\pi g^{2}\rho_{0} & \text{if \ensuremath{\lambda\leq\omega}\ensuremath{\leq D+\lambda}}\\
0 & \text{otherwise}
\end{cases} 
\end{align}

Since $\lambda$ is positive, it implies that $\omega$ should also be positive. In the Matsubara frequency space, we have $\Sigma_{\xi}(\bm{k},i\omega_{n})$
\begin{align}
    \Sigma_{\xi}(\bm{k},\omega_{n}) 
    = &  \sum_{\bm{p}}\left(\frac{g^{2}}{i\omega_{n}-\lambda-\varepsilon_{\bm{p}}}\right)+\sum_{\bm{p}}^{\sim}\left(\frac{g^{2}}{-i\omega_{n}+\varepsilon_{\bm{p}}+\lambda}\right) \nn
	 = & g^{2}\rho_{0}\int_{-D}^{D}\frac{d\varepsilon}{i\omega_{n}-\lambda-\varepsilon}+g^{2}\rho_{0}\int_{-D}^{0}\frac{d\varepsilon}{-i\omega_{n}+\varepsilon+\lambda} \nn 
	= & g^{2}\rho_{0}\ln\left(\frac{i\omega_{n}-\lambda+D}{i\omega_{n}-\lambda-D}\right)+g^{2}\rho_{0}\ln\left(\frac{i\omega_{n}-\lambda}{i\omega_{n}-\lambda+D}\right) \nn
	 = & g^{2}\rho_{0}\ln\left(\frac{i\omega_{n}+D}{i\omega_{n}-D}\right)+g^{2}\rho_{0}\ln\left(\frac{i\omega_{n}-\lambda}{i\omega_{n}+D}\right) \nn
	 = & -i\pi g^{2}\rho_{0}\text{sgn}(\omega_{n})+\frac{1}{2}g^{2}\rho_{0}\ln\left(\frac{\omega_{n}^{2}+\lambda^{2}}{\omega_{n}^{2}+D^{2}}\right) 
	 +ig^{2}\rho_{0}\left[\theta_{1}(\omega_{n})-\theta_{2}(\omega_{n})\right],
	\end{align}
	where 
	\begin{align}
	    \theta_{1}(\omega_{n})	=\tan^{-1}\left(\frac{\omega_{n}}{-\lambda}\right), \,\,
\theta_{2}(\omega_{n})	=\tan^{-1}\left(\frac{\omega_{n}}{D}\right).
	\end{align}
Thus, we obtain the imaginary part of $\Sigma_{\xi}$ 
\begin{align}
    \Sigma_{\xi}^{\prime\prime}(\bm{k},\omega_{n})	
     =& -\pi g^{2}\rho_{0}\text{sgn}(\omega_{n})+g^{2}\rho_{0}\left[\tan^{-1}\left(\frac{\omega_{n}}{-\lambda}\right)-\tan^{-1}\left(\frac{\omega_{n}}{D}\right)\right] \nn
	 =& -\pi g^{2}\rho_{0}\text{sgn}(\omega_{n})-g^{2}\rho_{0}\left[\frac{\omega_{n}}{\lambda}+\frac{\omega_{n}}{D}\right] \nn
	 \approx & -\pi g^{2}\rho_{0}\text{sgn}(\omega_{n}) \; ,
\end{align}
where, in the last line of the above equation, we have assumed $\lambda, \, D\gg \omega_n$. Combing $\Sigma_{\xi}^{\prime}$ and $\Sigma_{\xi}^{\prime\prime}$, the dressed retarded Green's function of the $\xi$ field up to the leading non-trivial self-energy correction is given by 
\begin{align}
    \mathcal{G}_{\xi}^{R}(\omega+i\epsilon,\bm{k})	= & \frac{1}{-t-\Sigma_{\xi}^{\prime}(\omega+i\epsilon,\bm{k})-i\Sigma_{\xi}^{\prime\prime}(\omega+i\epsilon,\bm{k})} \nn
	\equiv & \frac{1}{\omega_{\xi}-\xi_{\bm{k}}+i\Gamma_{\xi}},\quad\text{with }\text{\ensuremath{\lambda}\ensuremath{\leq\omega}\ensuremath{\leq D},}
\end{align}
where the quasi-particle width $\Gamma_{\xi}$, the generated  frequency $\omega_{\xi}$, and the dispersion $\xi_{\bm{k}}$ are given by 
\begin{align}
    &  \Gamma_{\xi}\equiv\pi g^{2}\rho_{0}, \nn
    &  \omega_{\xi}\equiv\frac{g^{2}\rho_{0}}{D}\omega\equiv\zeta\omega, \nn
    & \xi_{\bm{k}}= -\zeta\varepsilon_{\bm{k}}+\zeta\lambda+t.
\end{align}
Here, $\zeta\equiv\frac{g^{2}\rho_{0}}{D}$, $\rho_{0}\sim1/D$ and $\varepsilon_{\bm{k}} = h_{\bm{k}} -\mu_f \approx v |\bm{k}| $ with $\mu_f \equiv \mu - \lambda $ being the effective chemical potential for $f$ spinon. Here, the $f$-spinon band is approximated by a linear-in-momentum dispersion near the effective chemical potential $\mu_f$. As we treat $g$ as a perturbation, we expect $\zeta=(g/D)^{2}\ll1$. In addition, as the $\xi$ fermion is dispersionless at the bare level, its bandwidth of the renormalized band structure falls in the range of  $-\zeta D \leq \zeta \varepsilon_{\bm{k}}\leq\zeta D$ for $\bm{k}$ to be satisfactory with $-D\leq\varepsilon_{\bm{k}}\leq D$. The value of $\zeta D=g^{2}/D\sim t\zeta\ll t$. This leads to $\xi_{\bm{k}}>0$.

Following Ref.~\cite{YYC-ROPP-2025}, we generate and estimate  the dispersion of the slave boson $b$. At the bare level, the slave boson $b$ shows  only flat band with an atomic energy level $\lambda$. Here, 
the effective hopping of slave boson can be obtained by contracting the $\xi$ and $f$ fermions in the second-order perturbation of the hopping term $H_t$ as follows
\begin{align}
     & g^2 \sum_{ i,j,\sigma}\sum_{ i^\prime,j^\prime,\sigma^\prime} \left\langle  f_{i\sigma}^{\dagger} \xi_{ij,\sigma}\xi^\dagger_{i^\prime j^\prime,\sigma^\prime}f_{i^\prime \sigma^\prime} \right\rangle b_{j}^{\dagger}b_{j^\prime} \nn
    \sim &   g^2 \sum_{ i,j,\sigma}\sum_{ i^\prime,j^\prime,\sigma^\prime} \left\langle  f_{i\sigma}^{\dagger} f_{i^\prime \sigma^\prime} \right\rangle \left\langle \xi_{ij,\sigma}\xi^\dagger_{i^\prime j^\prime,\sigma^\prime} \right\rangle b_{j}^{\dagger}b_{j^\prime}.
\end{align}
Taking $\left\langle  f_{i\sigma}^{\dagger} f_{i^\prime \sigma^\prime} \right\rangle \sim \chi $ and $\left\langle \xi_{ij,\sigma}\xi^\dagger_{i^\prime j^\prime,\sigma^\prime} \right\rangle  \sim g^2/D^2 $, the dispersion of the slave boson field thus take the form $\varepsilon_b (\bm{k}) = \lambda + t_b (\cos k_x +\cos k_y)$ with the effective hopping  $t_b \propto g^4 \chi = t^4 \delta^2 \chi$ (in unit of \textit{D}).  For small hole-doping $\delta$ and a weakly perturbative coupling $t$ where $t/D \approx 3/4$ with $D \approx 4J_H$, $t_b$ can be estimated as $t_b \sim t^4 \delta^2 \sim (3/4)^4 \delta^2 $ (in unit of \textit{D}). We can fix the value of $\lambda$ such that $D>\lambda \gg |t_b|$. Therefore, in the wide-band limit, the effecitve boson hopping $t_b$ term of the generated slave-boson band is negligible, resulting in 
\begin{align}
   G_b (\bm{k},k_n) \approx \mathcal{G}^{\rm{B}}_b (\bm{k},k_n) = \frac{1}{ik_n -\lambda}\;.
\end{align}

\section{Renormalization group (RG) analysis}\label{app:RG}

Following Ref.~\cite{YYC-ROPP-2025}, in this section, we provide details of the RG analysis of our large-$N$ multi-channel Kondo lattice model. The RG procedure is primarily based on Refs.~\cite{Yamamoto-RG-PRB, qimiao-prb-local-fluc,lijun-prb-loca-fluc-bose-fermi}. 

\subsection{Scaling dimension}

The RG procedure starts from the scaling of the wave vector $\bm{k}$ and the frequency $\omega$, where 
\begin{align}
    k_i^{\prime}=e^{\ell}k_i,\quad\omega^{\prime}=e^{z \ell}\omega,
\end{align}
where $\ell$ is called the scaling factor, $i$ above denotes the component of $\bm{k}$, and $z$ represents the dynamical exponent. The bare scaling dimension for wave vector $k_i$ is chosen to be $[k_i]=1$. To preserve the Luttinger's theorem, the Fermi velocity has to be kept scale invariant. Since the dispersion of the $f$ spinon band is linear-in-momentum, we thus have
\begin{align}
    z=[\omega]=[\varepsilon_{\bm{k}}]=1.
\end{align}
The scaling dimension for the measure $d^{d}k$ for fermions with scale-invariant Fermi velocity is given by \cite{Yamamoto-RG-PRB}
\begin{align}
    [d^{d}k]=1.
\end{align}

The scale invariance for the action of the kinetic term of the $f$ spinon, $S_{0}^{f}$, implies that the bare scaling dimension of the  $f$ spinon operator is
\begin{align}
    1+1+1+2[f_{k\sigma}]=0\to[f_{k\sigma}]=-\frac{3}{2},
\end{align}
where $k \equiv (\omega,\, \bm{k})$. Since the bare slave boson $b_k$ is local, its scaling dimension can be thus obtained as
\begin{align}
[b_k]=-\left(\frac{d+z}{2}\right).
\end{align}
The non-interacting part of the composite fermion field $\xi^a_{k\sigma}$ is given by
\begin{align}
    S_{0}^{\xi}\sim\sum_{\sigma a}\int d\omega d^d k \left(\omega_{\xi}-\xi_{\bm{k}}\right)\left( \xi^a_{k\sigma}\right)^{\dagger}\xi^a_{k\sigma}, 
\end{align}
where the renormalized frequency $\omega_{\xi}=\frac{g^{2}\rho_{0}}{D}\omega$ and renormalized dispersion $\xi_{\bm{k}}=-\frac{g^{2}\rho_{0}}{D}\varepsilon_{\bm{k}}+\rm{const.}$. While performing the scaling procedure for $\xi_{\bm{k}}$ and $\omega_{\xi}$, we only scale $\varepsilon_{\bm{k}}$ and $\omega$. The prefactor $\frac{g^{2}\rho_{0}}{D}$ is not involved in the  scaling. This is equivalent to fixing $\frac{g^{2}\rho_{0}}{D}$ at its fixed point value, i.e. $\frac{g^{2}\rho_{0}}{D}\to \frac{(g^*)^{2}\rho_{0}}{D} $. We thus have the bare scaling dimension of the $\xi_{k\sigma}$, 
\begin{align}
    [\xi^a_{k\sigma}]=-\frac{3}{2}. 
\end{align}
The scaling dimension for various couplings can be computed in a similar approach. The action for $H_{t}$, denoted as $S_{t}$, takes the following form, 
\begin{equation}
    S_{t}\sim \bar{g}\int d\omega d\nu\int d^{d}kd^{d}p\sum_{a\sigma}\left[f_{k\sigma}^{\dagger}b_{p}^{\dagger}\bar{\xi}_{k+p,\sigma}^{a}+H.c.\right]
    \label{eq:S_t-action}
\end{equation}
with $k=(\omega,\bm{k})$ and $p=(\nu,\bm{p})$. From Ref. \cite{Yamamoto-RG-PRB}, for a special case $z=1$, the calculation of the scaling dimension for a boson-fermion coupling term can be performed via the usual way without invoking the patch scheme. The scaling dimension of $\bar{g}$ can be obtained similarly by simple power-counting, $ 0=[\bar{g}]+2z+1+d+[f_{k\sigma}]+[\xi_{k\sigma}]+[b_k]$. Consequently, we obtain
 \begin{align}
    [\bar{g}]	=\frac{4-d-3z}{2}=-\frac{1}{2}. 
 \end{align}
Note that, in Eq. (\ref{eq:S_t-action}), $\bm{k}$ denote the wave vector $\bm{k}$ for the $f$ spinon while $\bm{p}$ for the slave boson. Therefore, by Ref. \cite{Yamamoto-RG-PRB}, we take $[d^dk] = 1$ and $[d^d p] = d$ in the above derivation for $[\bar{g}]$. 
When $d = 2$ and $z =1$, 
\begin{align}
[\bar{g}] = -\frac{d-z}{2},
\end{align}
which we use in the RG $\beta$ functions in the next section.

\subsection{Self-energy corrections}

The self-energy correction to the bare Green's function of a given field begins with the following relation between the dressed (renormalized) Green's function $G$ of that field 
\begin{align}
    G=\mathbb{G}\mathcal{G} 
\end{align}
with $\mathcal{G}$ being the bare Green's function and $\mathbb{G}$ being the corrections to $\mathcal{G}$ by perturbation expansion. From the Dyson equation, we can also show that $G^{-1}=\mathcal{G}^{-1}-\Sigma=\mathcal{G}^{-1}\left(1-\mathcal{G}\Sigma\right)$ with $\Sigma$ here being denoted as the self-energy correction, implying that $ \mathbb{G}=\frac{1}{1-\mathcal{G}\Sigma}$. For weak coupling theory, we can expand $\mathbb{G}$ as a power series, 
\begin{align}
    \mathbb{G}\approx1+\mathcal{G}\Sigma+O(\mathcal{G}\Sigma)^{2}.
\end{align}
Note that the self-energy correction $\Sigma_{\xi}^{g}$ for the $\xi$ field in the RG analysis requires an additional negative sign compared to $\Sigma_{\xi}^{g}$ from RPA. This difference arises from the minus sign  appearing in the $k$-dependent term in the dispersion $\xi_{\bm{k}}$ for the $\xi$ field computed from RPA. As a result, $\Sigma_{\xi}^{g}$ is changed to
\begin{align}
    \Sigma_{\xi}^{g} (\omega_n) & = \left(\frac{\bar{g}^{2}}{NN_{s}\beta}\right)\sum_{\bm{p},p_{m}}\frac{1}{ip_{m}-\varepsilon_{\bm{p}}}\cdot\frac{1}{i\omega_{n}-ip_{m}-\lambda} \nn
    & = \left(-\frac{\bar{g}^{2}}{NN_{s}}\right)\sum_{\bm{p}}\left[\frac{n_{F}(\varepsilon_{\bm{p}})}{-i\omega_{n}+\varepsilon_{\bm{p}}+\lambda}+\frac{n_{F}(i\omega_{n}-\lambda)}{i\omega_{n}-\lambda-\varepsilon_{\bm{p}}}\right].
    \label{eq:Sigma_xi_RG}
\end{align}
Taking $-i\omega_{n}=-\omega-\lambda$, we have
\begin{align}
    \Sigma_{\xi}^{g} (\omega) & = 
    \left(-\frac{\bar{g}^{2}\rho_{0}}{N}\right)\int_{-D}^{0}\left(\frac{d\varepsilon}{-\omega+\varepsilon}\right)+\text{others} \nn
   & = \frac{\bar{g}^{2}\rho_{0}}{N} \ln\frac{D}{\omega}+\text{others},
\end{align}
 where ``others'' in the above equation represents the terms with no logarithmic correction.

 The self-energy of the $f$ spinon, $\Sigma_{f}^{g}$,  contributed from perturbation of $H_t$ is given by
\begin{align}
    \Sigma_{f}^{g} (\bm{k}, \omega_n) & = \left(-\frac{\bar{g}^{2}}{N_{s}\beta}\right) \sum_{p}\overline{\mathcal{G}}_{\xi}\left(k+p\right)\mathcal{G}_{b}\left( p \right) \nn
    & =\left(-\frac{\bar{g}^{2}}{N_{s}\beta}\right)\sum_{p_{m},\bm{p}}\frac{1}{i\omega_{n}+ip_{m}-\zeta^{-1}\xi_{\bm{k}+\bm{p}}}\cdot\frac{1}{ip_{m}-\lambda} \nn
    & = \frac{\bar{g}^{2}}{N_{s}}\sum_{\bm{p}}\frac{n_{F}(\zeta^{-1}\xi_{\bm{p}})}{i\omega_{n}+\varepsilon_{\bm{p}}+\zeta^{-1}\varUpsilon+\lambda},
\end{align}
where $\varUpsilon\equiv-\zeta\lambda-t+\mu_{\xi}$. 
Here, for the convenience of our RG analysis, we ignore the imaginary part $-i\pi D\sgn(\omega_n + p_m)$ in the denominator of the Green's function $\mathcal{G}_{\xi,\sigma a}(k)$. We checked that this term will not change the results of our one-loop RG analysis. 
After substituting $i\omega_{n} + \lambda \to \omega$, we obtain
\begin{align}
    \Sigma_{f}^{g} (\omega) = & \bar{g}^{2}\rho_{0}\int_{-\zeta^{-1}\varUpsilon}^{D}\frac{d\varepsilon}{\omega+\varepsilon} \nn
     = & \bar{g}^{2}\rho_{0}\ln\left(\frac{D}{\omega-\zeta^{-1}\varUpsilon}\right).
\end{align}
From the result in RPA section, we know that the self-energy of the slave boson contributes no logarithmic divergence, $\Sigma^g_b \propto (\ln D)^0$.

\subsection{Vertex corrections}

Vertex correction is defined as the correction for the bare vertex function due to interactions. If we denote the full vertex function as $\Pi$ and the bare one as $\Pi^{0}$. The vertex correction, represented as $U$, is defined through
\begin{align}
    \Pi = U \Pi^{0}.
    \label{eq:def-vertex-correction}
\end{align}
While employing perturbation expansion, $U$ can be written as $U=1+U^{(1)}+U^{(2)}+U^{(3)}+\cdots$ with $U^{(n)}$ being the vertex correction from the $n$-th order perturbation.

It is straightforward to show that there is no vertex corrections up to one-loop order contributed from the hopping  term $H_{t}$. 

In summary, the self-energy corrections of various fields from $H_{t}$ are given by
\begin{align}
    \Sigma_{\xi}^{g}(\omega) & \sim \frac{\bar{g}^{2}\rho_{0}}{N} \ln\frac{D}{\omega}, \nn
\Sigma_{b}^{g}(\omega)	& \sim(\ln D)^{0}, \nn
\Sigma_{f}^{g}(\omega)	& \sim\bar{g}^{2}\rho_{0}\ln\left(\frac{D}{\omega-\zeta^{-1}\varUpsilon}\right).
\end{align}
This leads to the following form of field renormalization $\mathbb{G}$:
\begin{align}
    \mathbb{G}_{f}	& \approx 1+ \left( \bar{g}^{2}\rho_{0}\right)\mathcal{G}_{f}\ln\left(\frac{D}{\omega-\zeta^{-1} \varUpsilon} \right) 
      \sim  1+ \left(\bar{g}^{2}\rho_{0}\right)\frac{1}{\omega}\ln\left(\frac{D}{\omega-\zeta^{-1}\varUpsilon}\right), \nn
\mathbb{G}_{\xi} & \approx 1+ \bar{g}^{2}\rho_{0} \mathcal{\bar{G}}_{\xi}\ln\frac{D}{\omega} \sim 1+\bar{g}^{2}\rho_{0}\frac{1}{\omega}\ln\frac{D}{\omega},\nn
\mathbb{G}_{b} & = 1. 
\end{align}
Here, we have replaced the bare Green's function with $\mathcal{G}\sim1/\omega$. In addition, we find the perturbative expansion to the vertex function of $H_t$ contributes the following vertex correction
\begin{align}
     U_{g}(k,p)	& =1. 
\end{align}

\subsection{The renormalization $z$-factors  and the RG $\beta$ functions}

To compute the RG $\beta$ functions, we start by reducing the cutoff energy from $D\to D^{\prime}=yD$ with $0<y<1$. This generates the relations between the wave-function (Green's function) and vertex corrections between $D$ and $D^\prime$, which shows that \cite{qimiao-prb-local-fluc,lijun-prb-loca-fluc-bose-fermi}
\begin{align}
    \mathbb{G}(\omega,D^{\prime},J_{i}^{\prime}) & =z_{\psi}\text{\ensuremath{\mathbb{G}}}(\omega,D,J_{i}),\nn
	U(\omega,D^{\prime},J_{i}^{\prime}) & =z_{U}^{-1}U(\omega,D,J_{i})
    \label{eq:G-U-renormal}
\end{align}
with $z_\psi$ and $z_U$ being denoted as the renormalization factor for the Green's function for the $\psi$ field and vertex correction $U$, respectively. The renormalization factor for the $\xi$ field is given by
\begin{align}
    z_{\xi}	& = \left(1+\frac{\bar{g}^{2}\rho_{0}}{N}\frac{1}{\omega}\ln\frac{D}{\omega}\right)^{-1}\left(1+\frac{\bar{g}^{2}\rho_{0}}{N} \frac{1}{\omega}\ln\frac{D^{\prime}}{\omega}\right) \nn
	        & =1-\frac{\bar{g}^{2}\rho_{0}}{N\omega}\ln \ell, 
\end{align}
where $\ln \ell \equiv\ln\frac{D}{D^{\prime}}$. As $\omega$ does not undergo a RG scaling [Eq. (\ref{eq:G-U-renormal})], we can set, for convenience, $\omega=D\to1/\omega=1/D=\rho_{0}$. This yields 
\begin{align}
    z_{\xi}=1-\frac{\bar{g}^{2}\rho_{0}^{2}}{N}\ln \ell 
    \to 1\quad(N \to \infty). 
\end{align}
Following a similar approach, we obtain the other two $z$ factors of fields,
\begin{align}
    z_{f} & =1-\left(\bar{g}\rho_{0}\right)^{2}\ln \ell.    
\end{align}
The $z$ factor for the coupling in $g$ is 
\begin{align}
    & z_g =1.  
\end{align}

\subsection{The RG scaling equations ($\beta$ functions)}

The renormalized couplings constants are defined as
\begin{align}
    \bar{g}^{\prime}&=z_{f}^{-\frac{1}{2}}z_{b}^{-\frac{1}{2}}z_{\xi}^{-\frac{1} {2}}z_{g}\bar{g}.  
\end{align}
Expanding $\bar{g}^\prime$ to the leading order in $\bar{g}\rho_0$, we obtain
\begin{align}
    \bar{g}^{\prime}&	=\left(1-\left(\bar{g}\rho_{0}\right)^{2}\ln \ell\right)^{-\frac{1}{2}}
    \bar{g} \nn
	 & =\left(1+\frac{1}{2}\left(\bar{g}\rho_{0}\right)^{2}\ln \ell\right)
     \bar{g} \nn
	& =\bar{g}+\kappa\bar{g}\left(\bar{g}\rho_{0}\right)^{2}\ln \ell. 
\end{align}
Multiplying both sides of $\bar{g}^{\prime}$ by $\rho_{0}$, we have 
\begin{align}
    \left({\bar{g}}\rho_{0}\right)^{\prime}	& =\bar{g}\rho_{0}+\kappa\left(\bar{g}\rho_{0}\right)^{3}\ln \ell.  
\end{align}
With the inclusion of the bare scaling dimension of $J$ and $\bar{g}$, these lead to the following RG $\beta$ functions:
\begin{align}
    \frac{d(\bar{g}\rho_{0})}{d {\ell}}	 & =-\left(\frac{d-z}{2}\right)(\bar{g}\rho_{0}) + \kappa\left(\bar{g}\rho_{0}\right)^{3}.  
\end{align}
The critical fixed point is at $(\bar{g} \rho_0)^2 = \epsilon/(2\kappa)$
where $\epsilon \equiv d-z$. In our case, we set $d=2, \, z=1$ and thus  $\epsilon=1$. Note that the RG scheme is still reliable for $\epsilon=1$ though it makes the $\epsilon$ expansion divergent \cite{goldenfeld-book-RG}.

\section{Demonstration of $\Sigma_b(\omega) \to 0$}\label{localbf}

The self-energy of the slave-boson $b$-field is defined as 
\begin{align}
    \Sigma_b(\omega_n) =& g^2 \sum_{\nu_n} G_{\xi}(\nu_n) G_{f}(\omega_n+\nu_n), \\
\mbox{where } G_{\xi}(\nu_n) =& \frac{1}{g^2} \sum_k 
            \frac{1}{i\nu_n-\epsilon_k-i\Gamma_{\xi}\sgn(\nu_n)} \\
\mbox{ with } \Gamma_{\xi} =& \pi D \text{ and } \nu_n = 2\pi n/\beta\,(n=\text{odd numbers}), \nonumber\\
            \text{and } G_{f}(\omega_n+\nu_n) =& \sum_k 
            \frac{1}{i(\omega_n+\nu_n)-\epsilon_k-\Sigma_f(\omega_n+\nu_n)}, \\
            \Sigma_f^{\prime\prime}(\omega_n) =& -\sgn(\omega_n) (\alpha-\zeta\omega_n) 
	\mbox{ with } \alpha = 2\kappa D/3 \text{ and } \zeta=2\kappa/\pi.\nonumber
\end{align}

Summing over momentum $k$ in Eq.~(S50) (Appendix E3 of \cite{YYC-ROPP-2025}), we obtain 
\begin{align}   
    G_f(i \omega_n) =& i 5\kappa \rho_0 -\frac{2\rho_0}{D} \Sigma_f^{\prime}(i \omega_n) 
                +i\frac{4\kappa \rho_0}{\pi D} |\omega_n| +\frac{2\rho_0}{D} i \omega_n, \nn
    g^2 G_{\xi}(i \omega_n) =& i 5\kappa \rho_0 
                +i\frac{4\kappa \rho_0}{D} \Gamma_{\xi} |\omega_n| +\frac{2\rho_0}{D} i \omega_n,
\end{align}
where $\Sigma_f^{\prime}(\omega_n) = \bar{g}^{2} \rho_{0} \ln\left(\frac{D}{\omega_{n}-\zeta^{-1}\varUpsilon}\right)$ and $\rho_0 = {1}/{D}$. 
Taking the analytic continuation, $i \omega_n \to z$ in Eq.~(S51), both Green's functions $G_f(z)$ and $G_{\xi}(z)$ do not exhibit poles in the complex $z$-plane. 
Therefore, $\Sigma_b = \frac{1}{g^2} \frac{1}{\beta} \oint \frac{dz}{2\pi} n(z) G_{\xi}(z) G_{f}(z+i\omega_n) \to 0$, 
where $n(z) = 1/[e^{-\beta z}+1]$ is the Fermi function. 
Finally, we obtain 
\begin{align}
    \Sigma_b(\omega) =& g^2 \sum_{\nu_n \neq 0} G_{\xi}(\nu_n) G_{f}(\omega_n+\nu_n) \to 0,\label{eq:Simga_b_0}
\end{align}
implying that the full Green's function of slave-boson $b$-field does not receive self-energy correction in the large-$N$ limit, and reduces to its bare form, $G_b=\mathcal{G}_b^B$.

\section{The large-$N$ multi-channel HFSBtJ formulation}\label{SBtJfor}

The large-$N$ multi-channel HFSBtJ formulation satisfy the following Dyson equations for the self-energy and dressed Green's functions (Eq. (3) and Eq. (4) of the main text),
\begin{align}
    \left[G_{\xi,\sigma a}(\omega)\right]^{-1} &=  \left[\mathcal{G}_{\xi,\sigma a}(\omega)\right]^{-1} - \Sigma_{\xi,\sigma a} (\omega) \; ,  \nn
    \left[G_{b,a}(\omega)\right]^{-1} & =  \left[\mathcal{G}_{b, a} (\omega)\right]^{-1}- \Sigma_{b,a}(\omega) \; , \nn
    \left[G_{f,\sigma}(\omega)\right]^{-1} & =  \left[\mathcal{G}_{f,\sigma}(\omega) \right]^{-1}- \Sigma_{f,\sigma}(\omega)\; ,  
\end{align}
and
\begin{align}
    \Sigma_{\xi,\sigma a} (\omega) & = - \frac{1}{N} \sum_{\nu} G_{f,\sigma} (\nu) G_{b,a} (\omega-\nu) \;, \nn
    \Sigma_{b,a} (\omega)  &= 2{g}^{2} \sum_{\nu,\sigma}  G_{f,\sigma}(\nu) G_{\xi,\sigma a} (\omega+\nu) \;, \nn
    \Sigma_{f,\sigma} (\omega)  &=  -\kappa g^2 \sum_{\nu,a} G_{\xi,\sigma a} (\omega+\nu) G_{b,a} (\nu)\; .
\end{align}
In the $N\to\infty$  limit,  $\Sigma_{\xi}(\omega) \propto O(1/N)$ is less relevant; and, as shown in the previous section, $\Sigma_b \approx 0$ in the wideband limit. As a result, we replace $G_{\xi}(\omega)$ and $G_{b}(\omega)$ by the ``initially RPA modified" one, i.e. $G _{\xi}(\omega)=\mathcal{G} _{\xi}(\omega)$ and $G _{b}(\omega)=\mathcal{G} _{b}(\omega)$. Thus, we directly compute $\Sigma _f$  using  $\mathcal{G} _\xi$ and $\mathcal{G} _b$, and the result is thus exact: namely $\Sigma _{f}(\omega)	=-\kappa g^{2} \sum_{\nu}\mathcal{G} _{\xi}(\omega+\nu)\mathcal{G} _{b}(\nu)$. The explicit form of $\Sigma _f$ can be straightforwardly calculated as 
\begin{align}
    \Sigma _{f}(\omega) 
    = & -\kappa g^{2} \sum_{p_n,\bm{p}} \frac{1}{ip_\xi  -\xi_{\bm{p}}-i\Sigma_\xi^{\prime\prime}(ip_n)} \frac{1}{ip_n-ik_n-\lambda} \;,
\end{align}
where $p_\xi \equiv \zeta p_n$. The result of the self-energy of the $f$ spinon can be found as
\begin{align}
    \left[\Sigma _f (\omega)\right]^{\prime\prime} = -\kappa \textrm{sgn}(\omega)(\alpha -  \varsigma \omega)\;.
\end{align}
with $\alpha\approx 2\kappa D/3$ and $\varsigma = 2\kappa/\pi$, as well as its dressed Green's function,
\begin{align}
   G _{f}(\omega)	= \sum_{\bm{k}} \frac{1}{\omega-\varepsilon_{\bm{k}}-\Sigma _{f}(\omega)}\;. 
\end{align}

With the exact Green's function of $G _f (\omega)$ shown above, we can then evaluate the resistivity contributing from the self-energy of the $\xi$ field from the hopping term. Due to the exact solutions of $G _f$ and $\Sigma _f$ in the large-$N$ limit, we note that the resulting resistivity contributing from the self-energy of the $\xi$ field is also exact.  

Note that though $\Sigma_{\xi} \sim O(1/N)$ is negligible in Eqs. (3) and (4) in the main text, it contributes to the total scattering rate by summing over $T$-matrix for $N$-spin flavors. 
The electron scattering (relaxation) rate $1/\tau_{\rm re}$ is given by the imaginary part of the $T$-matrix: $1/\tau_{re}=1/\tau_c\equiv-NT^{\prime\prime}(\omega)$ with $T^{\prime\prime}(\omega)= -2\Sigma_c^{\prime\prime}(\omega)$, and $\Sigma_c^{\prime\prime}=\frac{1}{2}\Sigma_{\xi}^{\prime\prime}$. 
We therefore evaluate $N \Sigma_{\xi}$, given by
\begin{align}
    N \Sigma_{\xi}(ik_{n})	= & \frac{1}{N_{s}\beta} \sum_{p}G_{f}(p)\mathcal{G}_{b}(k-p) \nn
	= & \frac{1}{\beta} \sum_{ip_{m}}\left(\frac{1}{N_{s}}\sum_{\bm{p}}G_{f}(p)\right)\mathcal{G}_{b}(k-p). 
 \label{eq:Sigma_xi}
\end{align}
Due the local nature (momentum-independent) of the slave boson (so is its Green's function), $\Sigma_\xi$ can be found to be independent of momentum. We first evaluate the momentum-independent dressed Green's function of the $f$ spinon via integrating over the momentum, which we obtain (Eq.~(E46) in Appendix E3 in Ref.~\cite{YYC-ROPP-2025})
\begin{align}
    G_{f}(p_{m}) = &  i\pi \rho_{0}-\frac{2\rho_{0}}{D}\Sigma_{f}(p_{m})+\frac{2\rho_{0}}{D}ip_{m} \nn
    = &  i\left(\pi - \frac{2\kappa}{3}\right)\rho_{0}-\frac{2\rho_{0}}{D}\Sigma_{f}^{\prime}(p_{m}) 
     +\frac{2 \kappa \rho_{0}}{\pi D}\text{sgn(\ensuremath{p_{m}})}ip_{m}   +\frac{2\rho_{0}}{D}ip_{m}.
\end{align}
Plugging the above result of $G_f(p_m)$ back to Eq.~\eqref{eq:Sigma_xi}, we obtain the real part of $\Sigma_\xi$ as (Eq.~(E52) in Appendix E3 in Ref.~\cite{YYC-ROPP-2025}) 
\begin{align}
    N\Sigma_{\xi}^{\prime}(k_{n}) = & \frac{2\rho_{0}^{2}}{D}\ln\left(\frac{D}{k_{n}-\zeta^{-1}\varUpsilon}\right)
     - \frac{2\rho_{0}}{D}J^{2}+\frac{2\rho_{0}\lambda}{D} +\frac{2\rho_{0}\lambda}{\pi D}\text{sgn}(k_{n}) \nn
    & + \frac{2\rho_{0}}{\pi D}\text{Re}\left(\int_{-\infty}^{\infty}\frac{dx}{\pi i}\cdot\frac{xn_{F}(x)}{x-ik_{n}+\lambda}\right) \; ,
\end{align}
and imaginary part as (Eq.~(E53) in Appendix E3 in Ref.~\cite{YYC-ROPP-2025}) 
\begin{align}
  N\Sigma_{\xi}^{\prime\prime}(\omega+i\delta) 
  = & -\left(\pi - \frac{2\kappa}{3}\right)\rho_{0}-\frac{2\rho_{0}}{\pi^{2}D}\omega_{0}-\frac{4\kappa \rho_{0}}{\pi D}\left|\omega\right|
   - \frac{2\rho_{0}}{\pi^{2}D}\left(\omega-\lambda\right)\ln\frac{|\omega-\lambda|}{\omega_{0}}.
\end{align}
For $\lambda \gg \omega$, we may simply approximate $\ln\frac{|\omega-\lambda|}{\omega_{0}} \approx \ln\frac{\lambda}{\omega_{0}}$. In addition, the linear-in-$\omega$ term in the last term of $\Sigma_\xi^{\prime\prime}$ above can be shown to vanish when we calculate the electrical conductivity below (due to the symmetric boundary for the frequency integral). We therefore can neglect this term. Now,  $N\Sigma_{\xi}^{\prime\prime}(\omega+i\delta)$ becomes
\begin{align}
    N\Sigma_{\xi}^{\prime\prime}(\omega+i\delta) = -A - \varsigma |\omega|
    \label{eq:Im-Sigma-xi-final}
\end{align}
with $A = \left(\pi - \frac{2\kappa}{3}\right)\bar{g}^{2}\rho_{0} + \frac{2\rho_{0}\bar{g}^{2}}{\pi^{2}D}\omega_{0}-\frac{2\rho_{0}\bar{g}^{2}}{\pi^{2}D}\lambda \ln\frac{\lambda}{\omega_{0}} $ and $
\varsigma = \frac{4\kappa}{\pi}$ a constant here.

\subsection{Scaling of AC resistivity}

We now extend the $N\Sigma_{\xi}^{\prime\prime}(\omega+i\delta) = -A - \varsigma |\omega|$ which is obtained at zero temperature to the finite-temperature regime via conformal transformation assuming the system is in the quantum critical scaling region. Then we 
use the finite-temperature result of $N\Sigma_{\xi}^{\prime\prime}(\omega,T)=2N\Sigma_c^{\prime\prime} (\omega,T)$ to evaluate the AC resistivity.

We start from the equation below for the spectral representation of correlation function \cite{georges-multichannel-kondo-PRB},
\begin{align}
    G_{\psi}(\tau)=-\int_{-\infty}^{\infty}\frac{e^{-\tau \varepsilon}}{1+e^{-\beta\varepsilon}}A_{\psi}(\varepsilon)d\varepsilon,
\label{eq:seeback-spectral}
\end{align}
where $0\leq\tau\le\beta$ and $A_\psi (\varepsilon) = (-1/\pi) G^{\prime\prime}_\psi( \omega+i0^+)$ denotes the spectral function for the $\psi$ field. When a system has conformal invariance, any fermionic correlation function, such as the self-energy, in the imaginary-time domain can be expressed as \cite{Georges-2021-PRR-seeback}
\begin{align}
    \Sigma(\tau)\propto e{}^{\alpha(\tau/\beta-1/2)}\left[\frac{\pi/\beta}{\sin\left(\pi\tau/\beta\right)}\right]^{1+\nu},
    \label{eq:seeback-self-energy-1}
\end{align}
where $\alpha$ and $\nu$ are constants. Here, $\alpha$ is a measure of particle-hole asymmetry. Its spectral representation is given by
\begin{align}
      e^{\alpha(\tau/\beta-1/2)}\left[\frac{\pi}{\sin\left(\pi\tau/\beta\right)}\right]^{1+\nu} 
     = C_{\alpha,\nu}\int_{-\infty}^{\infty}dx\frac{e^{-x\tau/\beta}}{1+e^{-x}}g_{\alpha,\nu}(x) 
     \label{eq:seeback-self-energy-2}
\end{align}
with
\begin{align}
    g_{\alpha,\nu}(x) & =\left|\Gamma\left(\frac{1+\nu}{2}+i\frac{x+\alpha}{2\pi}\right)\right|^2
      \frac{\cosh(x/2)}{\cosh(\alpha/2)\Gamma[(1+\nu)/2]^{2}}\, \nn
     C_{\alpha,\nu} & =\frac{(2\pi)^{\nu}\cosh(\alpha/2)\Gamma\left[\frac{1+\nu}{2}\right]^{2}}{\pi\Gamma[1+\nu]}\;.
     \label{eq:seeback-g}
\end{align}
Applying Eq. (\ref{eq:seeback-spectral}), we first evaluate the (imaginary) time dependence of a  spectral function which shows a linear-in-$|\varepsilon|$ dependence , $A_{\psi}(\varepsilon) = -|\varepsilon|/\pi$.  It can be shown that $A_{\psi}$ exhibits a $\tau^{-2}$ dependence. This  corresponds to the case of $\alpha = 0$ (particle-hole symmetric) and $\nu=1$ (Planckian) as discussed in Refs.~\cite{georges-multichannel-kondo-PRB,Georges-2021-PRR-seeback}. The second term of the above equation vanishes when $T \to 0$.

Now, we apply the above result to our theory. Once $\alpha=0$ and $\nu = 1$  are decided, we can  generalize $\Sigma_{\xi}$ which is initially obtained at zero temperature and shows a linear-in-frequency dependence, to the finite-temperature region by conformal transformation governed by Eqs. (\ref{eq:seeback-spectral})-(\ref{eq:seeback-g}). Replacing $\Sigma$, leading to the following expression for $N\Sigma_{\xi}^{\prime\prime}(\omega,T)$,
\begin{align}
    N\Sigma_{\xi}^{\prime\prime}(\omega,T)=\lambda_{0}\beta^{-1}g_{0,1}(x)=\frac{\lambda_{0}}{2}\omega\coth\left(\frac{\omega}{2T}\right)
    \label{eq:Sigma_xi-img-generalscaling}
\end{align}
with $\lambda_0$ being an unknown constant. $\lambda_0$ can be determined from its zero-temperature limit, where $N\Sigma_{\xi}^{\prime\prime}(\omega,T=0) = -(4 \kappa / \pi) |\omega| $. We thus find $\lambda_0 = -8 \kappa /\pi$, leading to 
\begin{align}
    N\Sigma_{c}^{\prime\prime}(\omega,T)=-\frac{2\kappa}{\pi}\omega\coth\left(\frac{\omega}{2T}\right)\;.
    \label{eq:Sigma_xi-img-generalscaling}
\end{align}
Using the relation  $\hbar/\tau_c = -2N\Sigma_c^{\prime\prime}$, we have
\begin{align}
    \frac{\hbar/\tau_c}{k_B T} = \frac{4\kappa}{\pi}x\coth \left(\frac{x}{2}\right)\;,
\end{align}
with $x\equiv \hbar \omega/k_B T$. In the $\omega/T \to \infty$ limit, scattering rate (divided by $k_B T$) exhibits a universal scaling behavior  $\frac{\hbar/\tau_c}{k_B T} = (4\kappa/\pi)x$. In contrast, in the DC-limit, the scattering rate shows the Planckian scattering rate, with a coupling-insensitive feature $1/\tau_c = \alpha_P k_B T$ with $\alpha_P = 8\kappa/\pi = 4/\pi \sim 1.27\,(\text{for }\kappa=1/2)$.

Note that the $\omega/T$-scaling function in scattering rate at finite frequency and temperature is not uniquely determined by the zero-temperature conduction electron self-energy 
$\Sigma_{\xi}^{\prime\prime}(\omega,T=0)$ via conformal mapping by Eqs.~(S65)-(S68). We find that in general the scaling function $\frac{\hbar/\tau}{k_BT} = \frac{2x}{\pi} \coth\frac{x}{2p}$ for $p \ge 1$ reduces to the same zero-temperature scattering rate $1/\tau(\omega,T=0) = -2\Sigma_{\xi}^{\prime\prime}(\omega,T=0) = \frac{2}{\pi} \omega$ in the limit of 
$\frac{k_BT}{\hbar\omega} \to 0$ (Fig.~S1). Here, we choose $p=1$ to reach a good agreement with the Planckian coefficient $\alpha \sim 1$ in LSCO observed in Ref.~\cite{Takagi_1992_LSCO_PRL} and Ref.~\cite{Ramshaw-MottPlanckian-arxiv-2024}. In Ref.~\cite{YYC-ROPP-2025}, however, we chose $p=2$ to get the best fit to the optical conductivity of LSCO observed in Ref.~\cite{George-NatComm-SM}.

\section{Estimation of $\rho_{\text{MIR}}$}\label{app:rMIR}

In this part, we estimate the in-plane $\rho_{\text{MIR}}$ based upon the Drude formula, given by
\begin{eqnarray}
\rho_{ab} &=& \frac{m^*}{ne^2} \frac{1}{\tau} d 
= \frac{2\pi m^*}{k_F^2 e^2} \frac{v_Fd}{v_F\tau} 
= \frac{2\pi m^*}{k_F^2 e^2} \frac{v_Fd}{l}.
\end{eqnarray}
The MIR limit for a strange metal is reached when $l \to 0$, leading to $\rho_{\text{MIR}}^{\text{P}} \propto \frac{v_F}{l} \to \infty$. However, for a normal metal, the MIR limit is accessible when the mean free path $l$ at least equals the Fermi wavelength $\lambda_F$, $l \ge \lambda_F = 2\pi/k_F$ or $k_F l \ge 2\pi$ \cite{Kivelson_1995_PRL_MIR}. Therefore, we have 
\begin{eqnarray}
\rho_{\text{MIR}} 
&=& \frac{2\pi m^*}{k_F e^2} \frac{v_Fd}{k_Fl} 
= \frac{2\pi m^*}{k_F e^2} \frac{v_Fd}{2\pi} 
= \frac{m^*v_Fd}{k_F e^2} \\
&=& \frac{m^*v_Fda}{\sqrt{2\pi (1-\delta)} e^2} 
\approx 0.444~\text{m}\Omega~\text{cm}.
\end{eqnarray}
In the last expression, we have used $k_F = \sqrt{2\pi n}$ with $d=0.64~\text{nm}$ \cite{Cooper_2009_Science_LSCO} being the interlayer spacing, $a=0.378~\text{nm}$ \cite{Cooper_2009_Science_LSCO} the intralayer lattice constant, $\delta=0.15$ \cite{Takagi_1992_LSCO_PRL} the hole-doped, $n=(1-\delta)/a^2 \approx 5.95~\text{nm}^{-2}$ the carrier density, $m^{\ast}=11.6~m_b$ ($m_b$ the bare electron mass) the effective mass, which agrees with the estimated value from \cite{Taillefer-planckian-2019} and with the low-$T$ static value of the effective mass \cite{George-NatComm-SM}. The Fermi velocity $v_F \approx 0.1 \times 10^6~\text{m}~\text{s}^{-1}$ is taken from Ref.~\cite{Gurvitch-PRL-1987}.

\section{Calculations of in-plane dc-resistivity and ac-conductivity of LSCO}\label{dc-ac}

In-plane dc-resistivity is calculated from the scattering rate predicted within our theory (Eq. (\ref{eq:uqc-scaling})) using the dc Drude formula $\rho_{ab} = (m^{\ast} d/ne^2)(1/\tau) = A_1 T$, where $A_1= (4/\pi) (k_B/\hbar) (m^{\ast}d/ne^2)$ is obtained by taking the dc limit of Eq. (\ref{eq:uqc-scaling}). The other parameters are $d=0.64~\text{nm}$ \cite{Cooper_2009_Science_LSCO} being the interlayer spacing, $a=0.378~\text{nm}$ \cite{Cooper_2009_Science_LSCO} the intralayer lattice constant, $\delta=0.15$ \cite{Takagi_1992_LSCO_PRL} the hole-doped, $n=(1-\delta)/a^2 \approx 5.95~\text{nm}^{-2}$ the carrier density, $m^{\ast}=11.6~m_b$ ($m_b$ the bare electron mass) the effective mass, which agrees with the estimated value from \cite{Taillefer-planckian-2019} and with the low-$T$ static value of the effective mass \cite{George-NatComm-SM}; these parameters have the same values as those used in SM \ref{app:rMIR} above.

In-plane ac-conductivity as a function of photon energy at different temperatures is computed at $\delta=0.24$ using the scattering rate given by Eq. (\ref{eq:uqc-scaling}). The mass enhancement is given by $\frac{m^{\ast}(\omega, T)}{m}-1\simeq -\frac{2}{\hbar\omega}\left[\Sigma_c^{\prime}\left(\frac{\hbar\omega}{2},T\right)-\Sigma_c^{\prime}(0,T)\right]$ with $m=2.76~m_b$ being the band mass \cite{George-NatComm-SM}. This implies the knowledge of the real part of the electrons' self energy, which can be usually obtained by using Kramers-Kroning relations $\Sigma^{\prime}_{c}(\omega, T)=\frac{1}{\pi}\mathcal{P}\int \frac{ \Sigma''_{c}(\omega_1)}{\omega_1-\omega} d \omega_1$. As explained in Ref. \cite{George-NatComm-SM}, considering a similar yet different scaling function $f(x)=|x|+2\exp{\frac{-|x|}{2}}$ [instead of $f(x)=x\coth(\frac{x}{2})$] with $x=\frac{\hbar\omega}{k_BT}$ allows to evaluate the real part of self energy. The resulting real part of self energy leads to a scaling of the mass enhancement $m^*/m-1 \propto g(x)$ with $g(x)$ taking the following form [Eq.~(S16) in Ref. \cite{George-NatComm-SM}] $g(x)= 2g\left( 1-\gamma \ln(\frac{x}{4})+\frac{2}{x}\left[ \exp\left(\frac{x}{4}\right)\textrm{Ei}(-\frac{x}{4}) \right. \right. \left. \left. -\exp\left(-\frac{x}{4}\right) \textrm{Ei}(\frac{x}{4})\right] \right)$. Finally, the in-plane optical conductivity is $\sigma_{ab}(\omega,T) = \sigma(\omega,T)/d$.






\begin{figure} 
	\centering
	\includegraphics[width=0.7\textwidth]{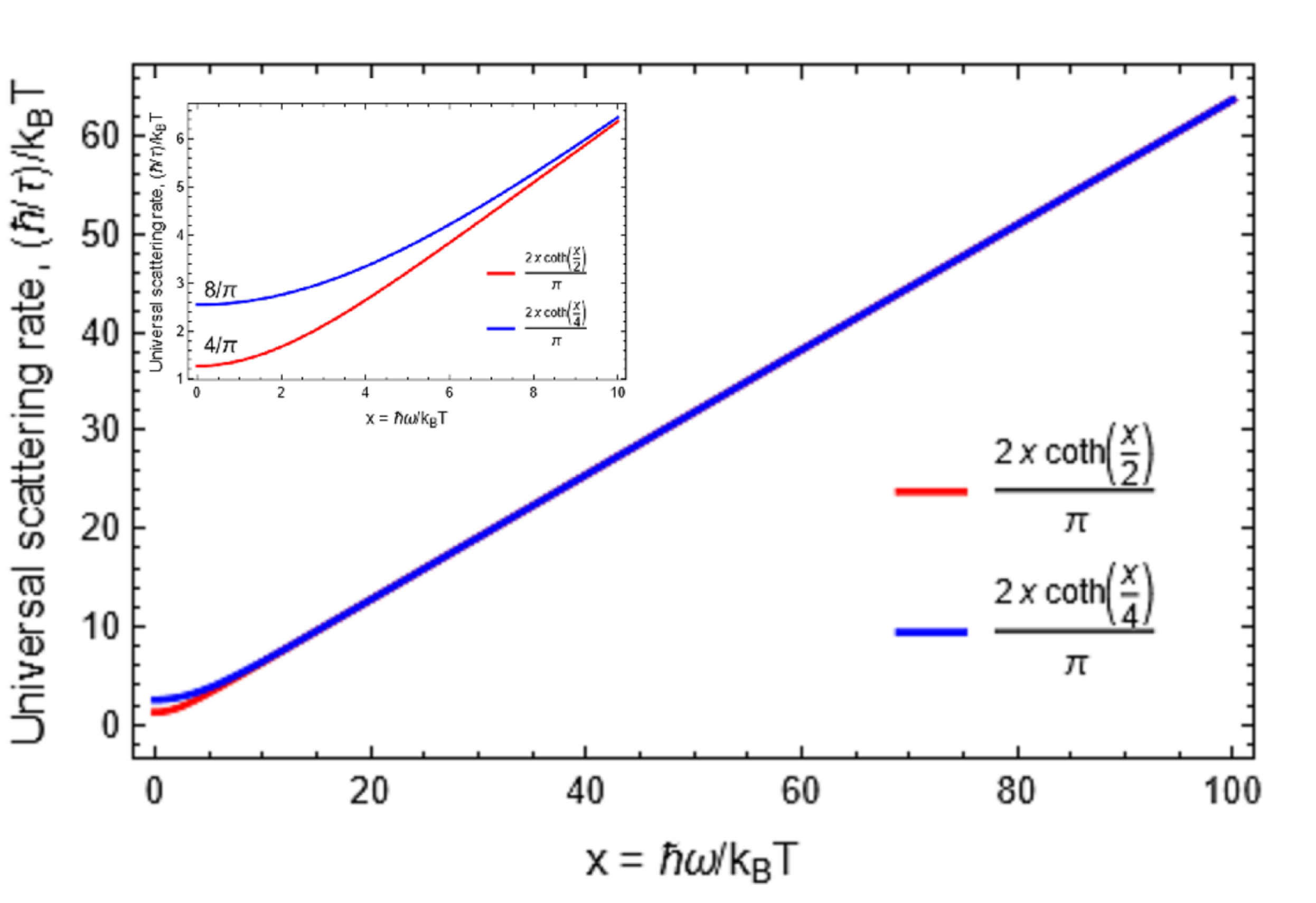} 

	\caption{\textbf{Universal scattering rate $\frac{\hbar/\tau}{k_BT} = \frac{2x}{\pi} \coth\frac{x}{2}$ (red) and $\frac{2x}{\pi} \coth\frac{x}{4}$ (blue).} 
	The inset figure is an enlarged part for $x \le 10$, which in the static limit approaches $\frac{4}{\pi}$ and $\frac{8}{\pi}$, respectively. 
	Note that the latter case, $\frac{\hbar/\tau}{k_BT} = \frac{2x}{\pi} \coth\frac{x}{4}$, has been considered in our recent publication \cite{YYC-ROPP-2025}.}
	\label{fig:usr} 
\end{figure}







\clearpage 





\end{document}